\documentclass[a4paper,fleqn,usenatbib]{mnras}

\usepackage{mathptmx}
\usepackage{txfonts}

\usepackage[T1]{fontenc}
\usepackage{ae,aecompl}

\usepackage{graphicx}
\usepackage{enumerate}
\usepackage{multirow}
\usepackage{xspace}
\usepackage{enumitem}
\usepackage[usenames, dvipsnames]{color}
\usepackage{soul}

\newcommand{\fof}{FoF\xspace}
\newcommand{\los}{LoS\xspace}
\newcommand{\com}{CoM\xspace}

\newcommand{\ssfr}{sSFR\xspace}
\newcommand{\umr}{$(u-r)_{corr}$}

\newcommand{\Efof}{E_{\mathrm{FoF}}}
\newcommand{\Emock}{E_{\mathrm{mock}}}
\newcommand{\Etot}{E_{\mathrm{tot}}}
\newcommand{\Qfof}{Q_{\mathrm{FoF}}}
\newcommand{\Qmock}{Q_{\mathrm{mock}}}
\newcommand{\Qtot}{Q_{\mathrm{tot}}}
\newcommand{\Stot}{S_{\mathrm{tot}}}

\newcommand{\Smock}{S_{\mathrm{mock}}}
\newcommand{\Sfof}{S_{\mathrm{\fof}}}

\newcommand{\Qfofi}{Q\ensuremath{_{\mathrm{FoF},i}}}
\newcommand{\Qmocki}{Q\ensuremath{_{\mathrm{mock},i}}}

\newcommand{\Pfofij}{P\ensuremath{_{\mathrm{FoF},ij}}}

\newcommand{\Pmockij}{P\ensuremath{_{\mathrm{mock},ij}}}
\newcommand{\Nmfofi}{N_{\mathrm{m}_{\mathrm{FoF}},i}}
\newcommand{\Nmfofj}{N_{\mathrm{m}_{\mathrm{FoF}},j}}
\newcommand{\Nmmocki}{N_{\mathrm{m}_{\mathrm{mock}},i}}
\newcommand{\Nmmockj}{N_{\mathrm{m}_{\mathrm{mock}},j}}

\newcommand{\Nmfof}{N_{\mathrm{m}_{\mathrm{FoF}}}}
\newcommand{\Nmmock}{N_{\mathrm{m}_{\mathrm{mock}}}}

\newcommand{\Nmfofbest}{N^{\mathrm{best}}_{\mathrm{m}_{\mathrm{FoF}}}}
\newcommand{\Nmfofbij}{N^{\mathrm{bij}}_{\mathrm{m}_{\mathrm{FoF}}}}

\newcommand{\Ngbij}{N_{\mathrm{g}_\mathrm{bij}}}
\newcommand{\Ngfof}{N_{\mathrm{g}_\mathrm{FoF}}}
\newcommand{\Ngmock}{N_{\mathrm{g}_\mathrm{mock}}}

\newcommand{\Nfof}{N_{\mathrm{FoF}}}

\newcommand{\cbest}{c^{\mathrm{best}}}
\newcommand{\cbij}{c^{\mathrm{bij}}}

\newcommand{\pbest}{p^{\mathrm{best}}}
\newcommand{\pbij}{p^{\mathrm{bij}}}

\newcommand{\Nbestgmock}{N^{\mathrm{best}}_{\mathrm{g}_\mathrm{mock}}}
\newcommand{\Nbijgmock}{N^{\mathrm{bij}}_{\mathrm{g}_\mathrm{mock}}}
\newcommand{\Nbestgfof}{N^{\mathrm{best}}_{\mathrm{g}_\mathrm{FoF}}}
\newcommand{\Nbijgfof}{N^{\mathrm{bij}}_{\mathrm{g}_\mathrm{FoF}}}

\newcommand{\Sgal}{S_{\mathrm{gal}}}
\newcommand{\finter}{f_{\mathrm{I}}}
\newcommand{\finterfield}{f_{\mathrm{I,field}}}

\newcommand{\Sgalbest}{S^{\mathrm{best}}_{\mathrm{gal}}}
\newcommand{\Sgalbij}{S^{\mathrm{bij}}_{\mathrm{gal}}}
\newcommand{\finterbest}{f^{\mathrm{best}}_{\mathrm{I}}}
\newcommand{\finterbij}{f^{\mathrm{bij}}_{\mathrm{I}}}

\newcommand{\Log}{\log}

\newcommand{\lgm}{\log\left(\frac{M_\star}{M_\odot}\right)}
\newcommand{\lgmsq}{\log^2\left(\frac{M_\star}{M_\odot}\right)}

\begin{document}

\title[Group quenching and galactic conformity at low $z$]{Group quenching and galactic conformity at low redshift}

\author[M. Treyer et al.]{M. Treyer,$^1$\thanks{E-mail: marie.treyer@lam.fr}
K. Kraljic,$^{1,2}$
S. Arnouts,$^1$
S. de la Torre,$^1$ 
C. Pichon,$^{3,4}$
Y. Dubois,$^3$
\newauthor
D. Vibert,$^1$
B. Milliard,$^1$
C. Laigle,$^5$
M. Seibert,$^6$
M. J. I. Brown,$^7$ 
M. W. Grootes,$^8$
 \newauthor
A. H. Wright,$^9$
J. Liske,$^{10}$
M. A. Lara-Lopez,$^{11}$
J. Bland-Hawthorn$^{12}$
\\  \\
$^1$ Aix Marseille Univ, CNRS, LAM, Laboratoire d'Astrophysique de Marseille, Marseille, France\\
$^2$ Institute for Astronomy, University of Edinburgh, Royal Observatory, Blackford Hill, Edinburgh, EH9 3HJ, UK \\
$^3$ Institut d'Astrophysique de Paris, UMR 7095 CNRS et Universit\'e Pierre et Marie Curie, 98bis Bd Arago, F-75014, Paris, France\\
$^4$ School of Physics, Korea Institute for Advanced Study (KIAS), 85 Hoegiro, Dongdaemun-gu, Seoul, 02455, Republic of Korea\\
$^5$ Department of Physics, University of Oxford, Keble Road, Oxford OX1 3RH, U.K.\\
$^6$ The Observatories of the Carnegie Institution for Science, 813 Santa Barbara St., Pasadena, CA 91101, USA\\
$^7$ School of Physics and Astronomy, Monash University, Clayton, Victoria, 3800, Australia\\
$^8$ ESA/ESTEC SCI-S, Keplerlaan 1, 2201 AZ, Noordwijk, The Netherlands\\
$^9$ Argelander-Institut f{\"u}r Astronomie, Universit{\"a}t Bonn, Auf dem H{\"u}gel 71, 53121 Bonn, Germany\\ 
$^{10}$ Hamburger Sternwarte, Universit{\"a}t Hamburg, Gojenbergsweg 112, 21029 Hamburg, Germany\\         
$^{11}$ Dark Cosmology Centre, Niels Bohr Institute, University of Copenhagen, Juliane Maries Vej 30, DK-2100 Copenhagen, Denmark\\
$^{12}$ Sydney Institute for Astronomy, School of Physics A28, University of Sydney, NSW 2006, Australia\\
}
\date{Accepted 2018 March 20. Received 2018 February 20; in original form 2017 September 18}
\pubyear{2018}

\maketitle

\begin{abstract}
We quantify the quenching impact of the group environment using the spectroscopic survey Galaxy and Mass Assembly (GAMA) to $z\sim0.2$. The fraction of red (quiescent) galaxies, whether in groups or isolated, increases with both stellar mass and large-scale (5 Mpc) density. 
At fixed stellar mass, the red fraction is on average higher for satellites of red centrals than of blue (star-forming) centrals, a galactic conformity effect that increases with density. Most of the signal originates from groups that have the highest stellar mass, reside in the densest environments, and have massive, red only centrals. Assuming a color-dependent halo-to-stellar-mass ratio, whereby red central galaxies inhabit significantly more massive halos than blue ones of the same stellar mass, two regimes emerge more distinctly: at $\Log(M_{halo}/M_\odot)\lesssim13$, central quenching is still ongoing, conformity is no longer existent, and satellites and group centrals exhibit the same quenching excess over field galaxies at all mass and density, in agreement with the concept of ``group quenching"; at $\Log(M_{h}/M_\odot)\gtrsim13$, a cutoff that sets apart massive ($\Log(M_{\star}/M_\odot)>11$), fully quenched group centrals, conformity is meaningless, and satellites undergo significantly more quenching than their counterparts in smaller halos. The latter effect strongly increases with density, giving rise to the density-dependent conformity signal when both regimes are mixed. The star-formation of blue satellites in massive halos is also suppressed compared to blue field galaxies, while blue group centrals and the majority of blue satellites, which reside in low mass halos, show no deviation from the color$-$stellar mass relation of blue field galaxies.
\end{abstract}


\section{Introduction}
\label{sec:introduction}

Most galaxies are either disky and actively forming stars, or spheroidal, with little or no ongoing star formation  \citep[e.g.][]{Strateva2001,Baldry2004}. This bimodality has been shown to exist up to redshift $z \sim 1$ \citep[e.g.][]{Bell2004,Tanaka2005,Willmer2006}, and possibly to $z \sim 2$ and beyond \citep[e.g.][]{Kriek2008,Brammer2009}. In between these so-called blue and red populations lies the little populated ``green valley" \citep{Wyder2007}, in which galaxies are thought to transit from star-forming to ``red and dead'' \citep{Krause2013}. The star-formation rate (SFR), which correlates with the density of gas in galaxy disks\citep{Schmidt1959,Kennicutt1998}, drops when the gas goes missing, a phenomenon known as quenching. Stellar mass and environment on various scales both seem to play a role in quenching galaxies \citep[e.g.][]{Peng2010}, but the physical processes most responsible for it remain elusive. 

The specific star-formation rate (\ssfr, the SFR per unit stellar mass) of blue galaxies is shown to decrease with increasing mass \citep[e.g.][]{Elbaz2007,Noeske2007,Salim2007},  the most massive galaxies being almost completely quenched. Several ``mass quenching'' mechanisms limited to massive halos have been proposed. Supplemented by virial shock heating of infalling cold gas \citep[][]{Birnboim2003, Keres2005, Dekel2006}, and gravitational heating due to clumpy accretion \citep[][]{Birnboim2007, Dekel2008, Dekel2009}, feedback from active galactic nuclei (AGN) has been found to be the most efficient mechanism to self-regulate the mass content of simulated massive galaxies \citep[e.g.][]{Croton2006, Hopkins2006, Sijacki2007} and the key ingredient to produce realistic mock massive galaxies in state-of the-art hydrodynamical cosmological simulations \citep[][]{Dubois2014, Dubois2016, Vogelsberger2014, Schaye2015}.
The stabilisation of gas disks \citep[``morphological quenching";][]{Martig2009} can explain quenching in less massive halos. 
 
The properties of galaxies with respect to their local environment -- on the scale of dark matter (DM) halos, i.e. groups and clusters -- have long been investigated, e.g. their color \citep[e.g.][]{Butcher1984,Balogh2004a}, morphology \citep[e.g.][]{Dressler1980,Dressler1997} or SFR \citep[e.g.][]{Gomez2003,Balogh2004b,Wijesinghe2012,Brough2013,Robotham2013,Davies2015}:  galaxies are known to be more often red and elliptical in clusters \citep[e.g.][]{Oemler1974,DavisGeller1976,Balogh1997,Balogh1999}.
 
Various local environmental quenching processes have been proposed. These include cluster-specific processes, such as galaxy harassment \citep{Moore1996}, ram pressure stripping of gas \citep{GunnGott1972} or interactions with the tidal field of clusters \citep{Byrd1990}, and group-specific processes, such as galaxy mergers \citep[e.g.][]{Toomre1972} and strangulation \citep{Larson1980,Balogh2000}. \citet{Prescott2011} found that the effect of environment on satellite galaxies was primarily a function of their host's stellar mass rather than their own stellar mass, and that strangulation was likely to be the main gas removal process that quenches them. \citet{Grootes2017} argued on the contrary that it was the ongoing substantial accretion of gas in groups that led to the buildup of spheroidal components in satellite disk galaxies, and eventually to their ``death by gluttony" rather than starvation. 

It is also debated whether it is the central-satellite dichotomy \citep[e.g.][]{Peng2012,Kovac2014} or simply being a member of a group that is crucial for the quenching of star formation, i.e. are we mostly seeing satellite quenching \citep[e.g.][]{Bosch2008,Hartley2015,Carollo2016} or group quenching \citep{Knobel2015}? 

In addition to local (halo scale) effects, the formation epoch and subsequent accretion history of a halo depend on its locus in the large-scale environment, a phenomenon referred to as ``assembly bias" \citep[e.g.][]{ShethTormen2004,Gao2005,Wechsler2006,Musso2017}. An observation, first made and coined ``galactic conformity" by \citet{Weinmann2006}, who analysed galaxy groups in the Sloan Digital Sky Survey \citep[SDSS;][]{sdss2000}, has been suggested as evidence of this past large scale environmental effect:  the fraction of quenched satellites around a quenched central is found to be significantly higher than around a star-forming central\footnote{\citet{Holmberg1958} first observed color conformity in galaxy pairs and concluded that it could not be accounted for by the known correlation with morphological type. For 20 years the Holmberg effect received attention from neither observers nor theoreticians, but it was later confirmed by several groups as evidence of coupled evolution (http://ned.ipac.caltech.edu/level5/Sept02/Keel/Keel5\_4.html).}, {\it at fixed halo mass}.  An interpretation might be that galaxy properties depend not only on the mass of their DM halos but also on their assembly history \citep[e.g.][]{Hearin2015}. Galactic conformity has since been confirmed by other analysis of the SDSS and observed at redshift $z\geq 2$ \citep{Hartley2015,Kawin2016}. It was also claimed to persist out to scales far larger than the virial radius of halos \citep{Kauffmann2013,Hearin2015}, but this puzzling result, named ``2-halo conformity", and which several studies have accounted for by advocating the mutual evolution of halos in the same large-scale tidal field \citep{Hearin2015,Hearin2016,ZuMandelbaum2017,RafieferantsoaDave2017}, was recently questioned and found attributable to methodological biases \citep{Sin2017}.


Quenching in groups has been studied \citep[e.g.][]{Bosch2008,Prescott2011,Peng2012,Wetzel2013,Knobel2015,Kawin2016} by means of red fractions (fractions of quiescent galaxies) and quenching efficiencies (excess quenching with respect to some control sample) as a function of various environmental parameters, such as halo mass, halo-centric distance, local galaxy density, or mass of the central galaxy. While there is general agreement on the dependence of satellite quenching on these parameters, a physical interpretation is not straightforward, mainly because the parameters are correlated and difficult to disentangle, even through multidimensional analysis \citep{Knobel2015}. 

The aim of this paper is to revisit the quenching impact of group environment, and to probe ``1-halo'' galactic conformity in particular, using the spectroscopic survey Galaxy and Mass Assembly \citep[GAMA; ][]{Driver2009,Driver2011,Liske2015}. Thanks to its depth and spectroscopic completeness, GAMA allows us to expand the SDSS investigations to $z\sim0.2$ in a significant volume of the Universe. From a group catalog we constructed using an anisotropic Friends-of-Friends algorithm taking into account the effects of redshift-space distortion, we study the red fraction, quenching efficiency and star-formation activity of galaxies in groups as a function of central galaxy color, group stellar mass, large-scale density, and finally halo mass. This group catalog, corrected for the finger-of-god effects, is used in a companion paper \citep{Kraljic2018} to improve the reconstruction of the cosmic web  and to explore the impact of its anisotropic features (nodes, filaments, walls, voids)  on galaxy properties.

The outline of the paper is as follows: in Section \ref{sec:data}, we describe the GAMA data, the derivation of the physical properties of the galaxies and our criterion for classifying them into star-forming and quiescent. We present the group catalog in Section \ref{sec:groups}. The stellar mass and environmental dependences of quenching and conformity are analysed in Section \ref{sec:results} and \ref{sec:environment} respectively. Star-formation in groups is explored in Section \ref{sec:sf}. We discuss the uncertainties of our analysis in Section \ref{sec:caveats} and conclude in Section \ref{sec:conclusions}. The paper also contains appendices dedicated to the detailed description of our group finder (adopted algorithm, mock catalogs used for its calibration, optimization strategy and tests of group reconstruction quality).

Throughout our analysis, we adopt a flat $\Lambda$CDM cosmology with H$_0 =$ 67.3 km s$^{-1}$ Mpc$^{-1}$, $\Omega_{m} = 0.3$ and $\Omega_{\Lambda} = 0.7$ 
 \defcitealias{Planck_cosmoparam2016}{Planck Collaboration 2016} \citepalias{Planck_cosmoparam2016}.
All magnitudes are quoted in the AB system and the physical parameters are derived assuming a Chabrier IMF \citep{Chabrier2003}. 

\section{Data}
\label{sec:data}

\begin{figure*}
\includegraphics[width=6.05cm]{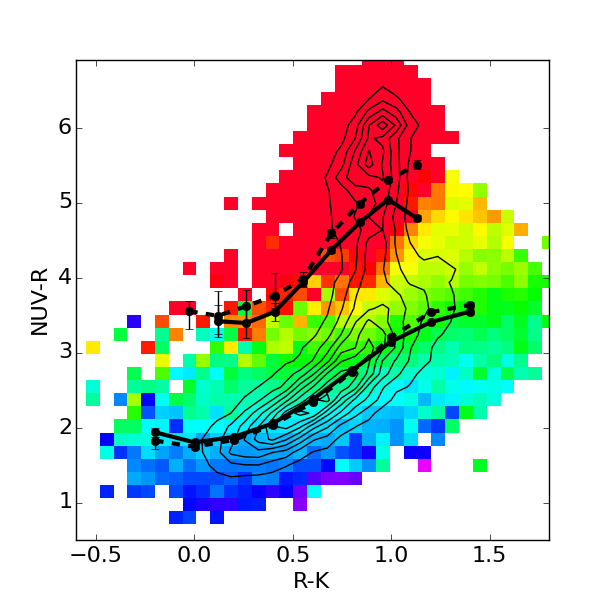}
\hspace{-0.5cm}
\includegraphics[width=11.9cm]{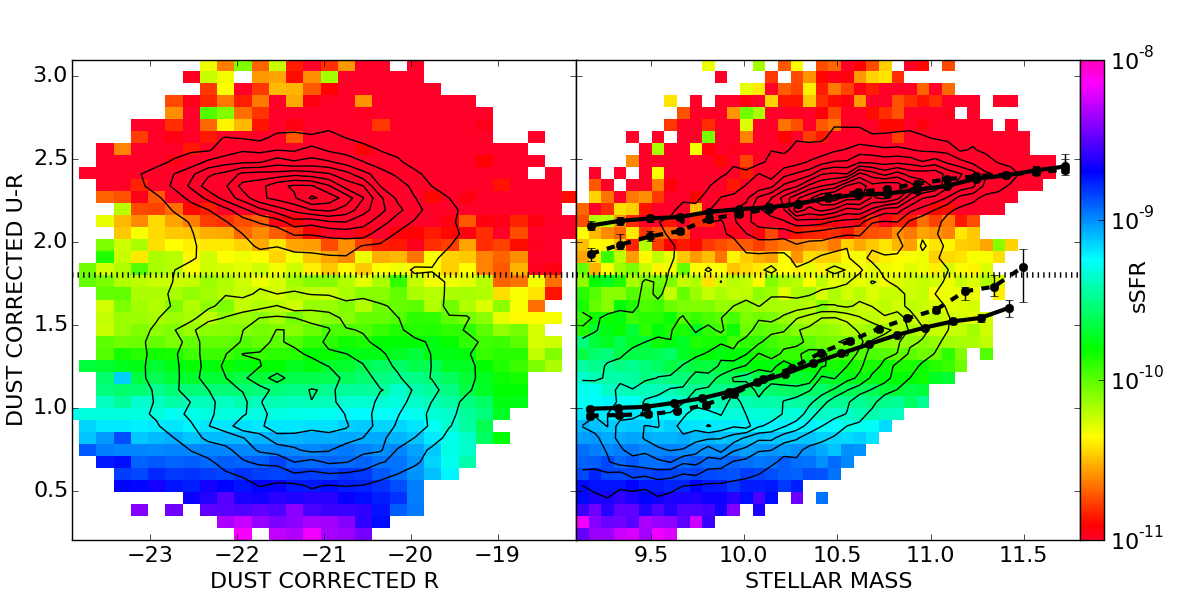}
\caption{\label{cmd} Rest frame $(NUV-r)$ versus $(r-k)$ colors (left panel), and dust corrected $(u - r)$ color versus $r$ magnitude (middle panel) and versus stellar mass (right panel) at $0.02 < z < 0.24$. Stellar masses are in units of $\Log (M_{\star} / M_\odot)$.The contours are unweighted number density contours. The color code reflects the average specific star-formation rate (\ssfr) per pixel. The straight color cut at $(u - r)_{corr} = 1.8$ (dotted line) is statistically consistent with a cut in \ssfr at $\sim 10^{-10.5}$ yr$^{-1}$ except for low mass red galaxies. The solid and dashed lines show the median relations for the quiescent and star-forming populations defined in terms of these \umr\ and \ssfr cuts. In the rest of this paper, the blue and red populations are defined in terms of dust corrected $(u-r)$ with a dividing line at $(u-r)_{corr}=1.8$.}
\end{figure*}

The GAMA\footnote{http://www.gama-survey.org/} survey \citep{Driver2009,Driver2011,Liske2015} is a joint European-Australian spectroscopic survey combining multi-wavelength photometric data from several ground and space-based programs. The photometric coverage includes data from the Galaxy Evolution Explorer (GALEX) in the far and near-ultraviolet (FUV and NUV), the Sloan Digital Sky Survey (SDSS) at optical wavelengths ($u$, $g$, $r$, $i$ and $z$ passbands), the VISTA Kilo-degree Infrared Galaxy (VIKING) Survey in the ZYJHK bands,
and the Wide-Field Infrared Survey Explorer (WISE) in four mid-infrared bands from 7 to 22 $\mu$m. Far-infrared (FIR) data from Herschel ATLAS (H-ATLAS) and radio data from the Giant Meterwave Radio Telescope have also been acquired. GAMA was intended to link wide and shallow surveys such as the SDSS Main Galaxy Sample \citep{Strauss2002} to narrow and deep surveys such as DEEP2 \citep{DEEP2}. 

The photometric data used in this work is the LAMBDAR\footnote{Lambda Adaptive Multi-Band Deblending Algorythm in R} panchromatic photometric catalog LambdarCatv01 \citep{Wright2016}, consisting of three equatorial fields (G09, G12, G15) covering a total of 180 deg$^2$ (3 times $12\times5$ deg$^2$). The spectroscopy was carried out using the 2dF/AAOmega multi-object spectrograph on the Anglo-Australian Telescope (AAT), building on previous spectroscopic surveys such as SDSS, the 2dF Galaxy Redshift Survey (2dFGRS) and the Millennium Galaxy Catalogue (MGC). 
It is nearly complete (98\%) to $r=19.8$ \citep{Liske2015}, each region of the sky being observed multiple times (the target density being much higher than the available fiber density), with at least one member of any given close-packed group receiving a fiber whenever that region was visited \citep{Robotham2010}. This makes GAMA a more suitable dataset to study galaxies in groups and close pairs than other spectroscopic surveys (e.g. SDSS) that miss a fraction of close targets, especially in high density regions (see \citet{Liske2015} for the GAMA completeness on small scales). 

\subsection{Physical parameters}
The physical quantities used in this work were derived from the spectral energy distribution (SED) fitting code LEPHARE\footnote{http://cesam.lam.fr/lephare/lephare.html} using the FUV to NIR photometry (11 bands). We used a set of model spectra from \citet{BruzualCharlot2003}, assuming a range of exponentially declining star-formation histories and a Chabrier IMF \citep{Chabrier2003}, as well as three dust obscuration laws: the commonly used starburst law of \citet{Calzetti2000}, an exponential law with exponent 0.9 and the Small Magellanic Cloud law of \citet{Prevot1984}. The physical quantities of interest in this paper are the stellar mass (defined as the median of the probability distributions), the dust-corrected absolute magnitudes, the specific star-formation rate (\ssfr) and the maximum volume $V_{max}$ in which a galaxy would remain observable above the survey flux limit given its luminosity and spectral type.

\subsection{Star-forming (blue) vs quiescent (red) classification}

While the bimodality in galaxy properties has been observed as far back as \citet{Hubble1926}, with late type, star-forming, spiral or irregular, blue galaxies on the one hand and early type, elliptical, ``red and dead" galaxies on the other, defining a transition between the two populations is not straightforward as the distributions overlap \citep{Taylor2015}. Color-magnitude or color-mass diagrams are most often used to draw the line, or a smooth transition zone, e.g. \citet{Taylor2015} introduced a statistical approach allowing for the natural overlap of the two populations in the ($g - i$) versus stellar mass diagram using GAMA at $z  < 0.12$. However shorter wavelengths prove most discriminating \citep[e.g. UV-optical colors;][]{Wyder2007}, even though dust causes confusion since dusty star-forming galaxies look red and may be mistaken for quiescent galaxies. 
This mixing can be efficiently sorted out by using color-color diagrams, such as  ($NUV-r$) vs $(r-k)$  which was shown to be a powerful diagnostic to separate dusty star-forming galaxies from intrinsically red, quiescent ones \citep{Arnouts2013}.
 
Figure \ref{cmd} shows the distribution of galaxies in the ($NUV-r$) versus $(r-k)$ diagram for the $\sim 85\%$ that are UV detected (left panel), and the dust corrected ($u - r$) versus $r$ and versus stellar mass diagrams (middle and right panels respectively), where $NUV$, $u$, $r$ and $k$ refer to the rest-frame magnitudes.
The color code reflects the average \ssfr per pixel. The dust corrected color and \ssfr are correlated through the modeling of the attenuation, but the two populations are separated equally well around the same \ssfr values in the uncorrected ($NUV-r$) vs $(r-k)$ diagram. 
 The consistency between the  dust uncorrected bi-color diagram and the dust corrected color gives us confidence in the dust recipes used in our SED fitting. 
 We find that a straight color cut at $(u - r)_{corr} = 1.8$ is consistent with a cut in \ssfr at $\sim 10^{-10.5}$ yr$^{-1}$. The solid and dashed lines show the median relations for the quiescent and star-forming populations defined in terms of these \umr\ and \ssfr boundaries, respectively. They form the well-known red and blue sequences. 

In the range of stellar masses that we use in this work ($\Log(M_{\star} /M_\odot) \gtrsim 10.25$, see next section), the above separation criteria yield undistinguishable results. Thus we simply define the blue and red populations in terms of dust corrected $(u-r)$ color with a dividing line at $(u-r)_{corr}=1.8$, and we will use the terms red (blue) and quiescent (star-forming) galaxies interchangeably. The term $(u-r)$, or U-R in figure labels, will always refer to the dust-corrected color.

\subsection{Stellar mass completeness}
\label{subsec:mass_completeness}

\begin{figure}
\begin{center}
\includegraphics[width=9.cm]{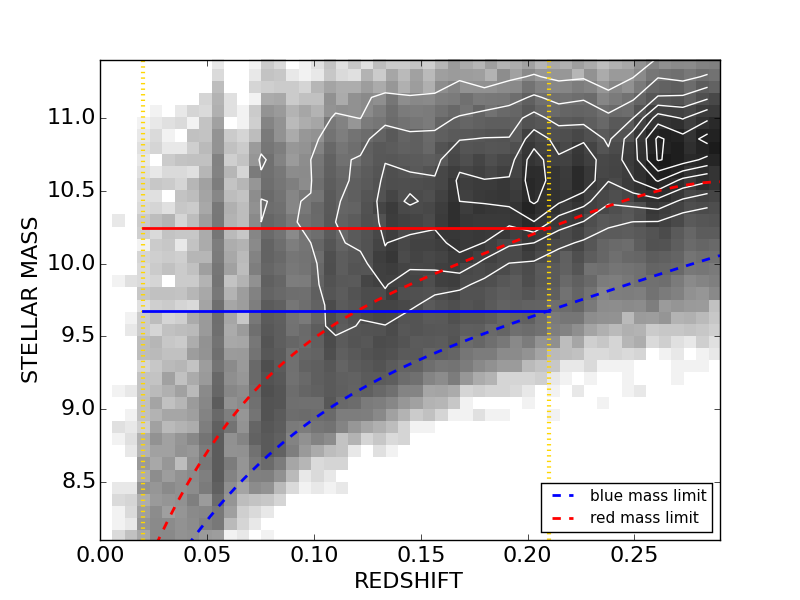}
\caption{The stellar mass in units of $\Log (M_{\star} / M_\odot)$ versus redshift distribution, with number density contours for the red population. The blue and red dashed lines represent the mass completeness limits for the blue and red galaxies respectively as a function of $z$. The vertical lines and the red horizontal line are the redshift and mass limits used in our analysis: $0.02<z<0.21$ and $\Log (M_\star/M_\odot)>10.25$}
\label{minmassz}
\end{center}
\end{figure}

In order to compute redshift and mass unbiased red fractions, we must restrict our sample to group members more massive than the completeness limit at the maximum redshift considered. Figure \ref{minmassz} shows the mass completeness limits as a function of redshift for the blue and red galaxies as blue and red dashed lines respectively. These limits are defined as the mass above which 90\% of the galaxies reside at a given redshift $z \pm 0.004$. The redshift/mass compromise used in the rest of this paper is $z<0.21$ and $\Log (M_\star/M_\odot)>10.25$ (the red mass limit in this redshift range).

\section{Group catalog}
\label{sec:groups}

\subsection{Group catalog construction}
\label{subsec:fofconstruction}

\begin{figure}
\begin{center}
\includegraphics[width=\columnwidth]{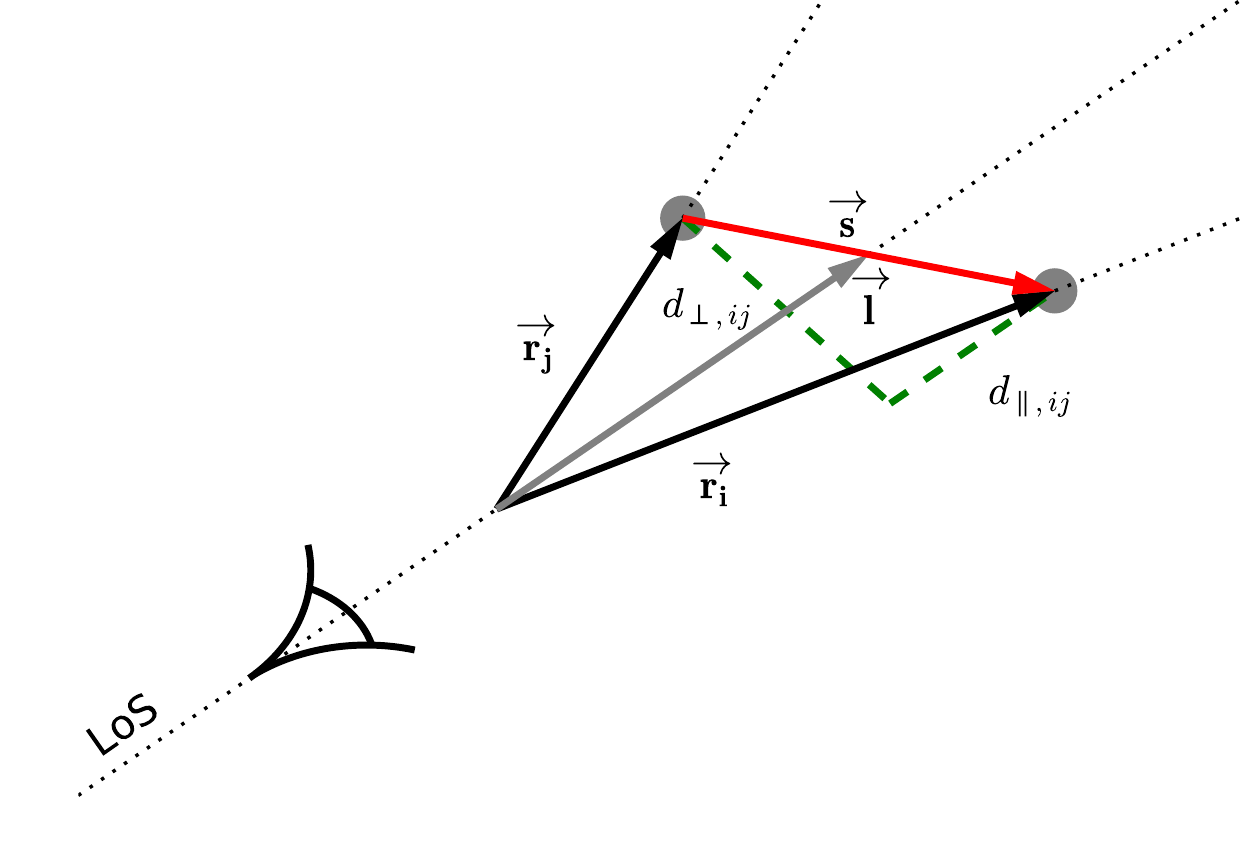}
\caption{\label{Fig:scheme} Schematic illustration of the definitions used in the \fof algorithm. In order to deal with the effects of redshift distortions, the separation $\vec{s}$ between two galaxies $i$ and $j$ at positions $\vec{r_i}$ and $\vec{r_i}$ is measured in two coordinates: the parallel $d_{\parallel,ij}$ (Eq.~\ref{eq:d_parallel}) and perpendicular $d_{\perp,ij}$ (Eq.~\ref{eq:d_perp}) projected comoving separations to the mean line-of-sight $\vec{l}$ (LoS; Eq.~\ref{eq:l}). }
\end{center}
\end{figure}

Although a GAMA group catalog already exists \citep{Robotham2011}, we developed our own tool for the purpose of an ongoing cosmic web study in this and other datasets \citep[e.g.][]{Malavasi2017,Kraljic2018}. 
A detailed description of the Friends-of-Friends (\fof) algorithm we adopted to detect the groups is presented in the appendices.  A schematic illustration of the method is depicted in Fig.~\ref{Fig:scheme}. In order to deal with the effects of redshift-space distortions, the distance between two galaxies $i$ and $j$ is measured in two coordinates: the parallel $d_{\parallel,ij}$ (Eq.~\ref{eq:d_parallel}) and perpendicular $d_{\perp,ij}$ (Eq.~\ref{eq:d_perp}) projected comoving separations to the mean line-of-sight $\vec{l}$ (Eq.~\ref{eq:l}). 
We next introduce two linking lengths $b_\perp$ and $b_\parallel$, the projected and line-of-sight linking lengths in units of  the mean intergalactic separation $\overline{r}_{ij}$ (Eq.~\ref{eq:mean_sep}), respectively, related through the radial expansion factor $R = b_\parallel / b_\perp$ accounting for the peculiar motions of galaxies within groups. Two galaxies are assumed to be linked to each other if their projected perpendicular and parallel separations are smaller than the corresponding linking lengths (Eq.~\ref{eq:link_perp} and Eq.~\ref{eq:link_parallel}).

The linking length $b_\perp$ and the radial expansion factor $R$ (or equivalently the perpendicular and line-of-sight linking lengths) are the two free parameters to be optimized.  Their values will affect the quality of the resulting group catalog: too small values will tend to break up single groups into several groups, while too large values will merge multiple groups into single ones. 

These free parameters can be determined from the optimization of some group cost function, depending on the scientific purpose of the group catalog, when tested on mock catalogs. Our objective is to obtain a catalog with a high group detection rate and a low contamination by galaxies coming from different groups. We followed the definition of the group cost function of \citet{Robotham2011}, $\Stot$, with slightly different notations and minor modifications. This cost function is meant to fulfil the requirement that the reconstructed and underlying real group catalog are mutually accurate representations of each other. By definition, $\Stot$ takes values between 0 and 1 and must be maximised. The method is described in Appendix \ref{appendix:optimization}. 

\begin{figure}
\begin{center}
\includegraphics[width=\columnwidth]{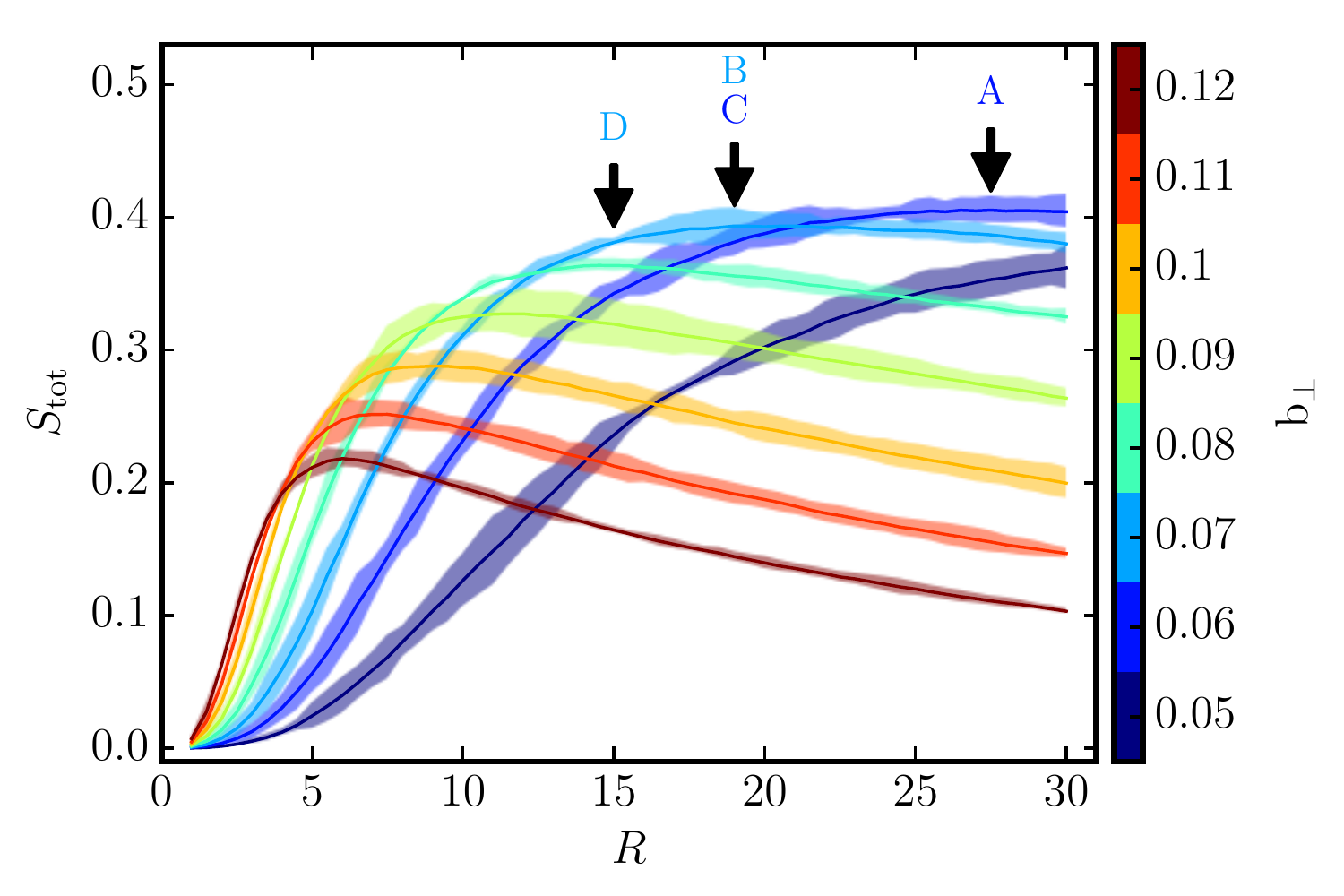}  
\caption{\label{Fig:stot} Group cost function $\Stot$ as a function of the radial expansion factor $R$ for different values of the linking length $b_\perp$.  The global maximum (A) is obtained for ($b_\perp$,$R$) = (0.06,27.5). However given the degeneracy between the two parameters, we include an additional criterion of symmetry between the recovered and real group and apply it to the following local maxima: ($b_\perp$, $R$) = (0.07, 19.0) corresponding to the maximum $\Stot$ for $b_\perp$=0.07 (B), ($b_\perp$, $R$) = (0.06, 19.0) corresponding to the most symmetric contribution from the mock and \fof groups to $\Stot$ for $b_\perp$=0.06 (C) and ($b_\perp$, $R$) = (0.07, 15.0) corresponding to the most symmetric solution for $b_\perp$=0.07 (D). Our final choice is (C) (see the appendices for more details).
}
\end{center}
\end{figure}

Results of the optimization computation are shown in Fig.~\ref{Fig:stot}. As can be seen, there is a degeneracy between the parameters $b_\perp$ and $R$: for values of $b_\perp$ equal to 0.06 and 0.07, $\Stot$ does not evolve significantly between $R \gtrsim 16$ and $R \simeq 30$, and between $R \gtrsim 14$ and $R \simeq 25$, respectively. This means that the global statistical properties of group catalogs constructed using a combination of $b_\perp$ and $R$ in these ranges will be similar.

Given this degeneracy, we include an additional criterion of symmetry between the recovered and real groups: we request that the individual contribution of the mock and \fof components to the overall cost function be similar (the more similar these contributions, the more similar the reconstructed groups of the mock and \fof catalogs). With this additional constraint, our final choice for the \fof parameters are $b_\perp$ = 0.06 and $R$ = 19.0. This combination of parameters is optimal when considering statistical measures of the group reconstruction quality independent of those used in the optimization, as shown in Appendix \ref{appendix:quality}. Our linking lengths are in good agreement with the combination found to be optimal for studies of environmental effects by \citet{DuarteMamon2014}, who tested the parameters according to the scientific goal of the group catalog. We did not apply any completeness correction to the linking parameters since the GAMA survey is spectroscopically extremely complete \citep[$\sim$ 98\% within the $r$-band limit,][]{Driver2011} and the mean modifications would be less than 1\% \citep{Robotham2011}.

\begin{figure*}
\includegraphics[width=8.7cm]{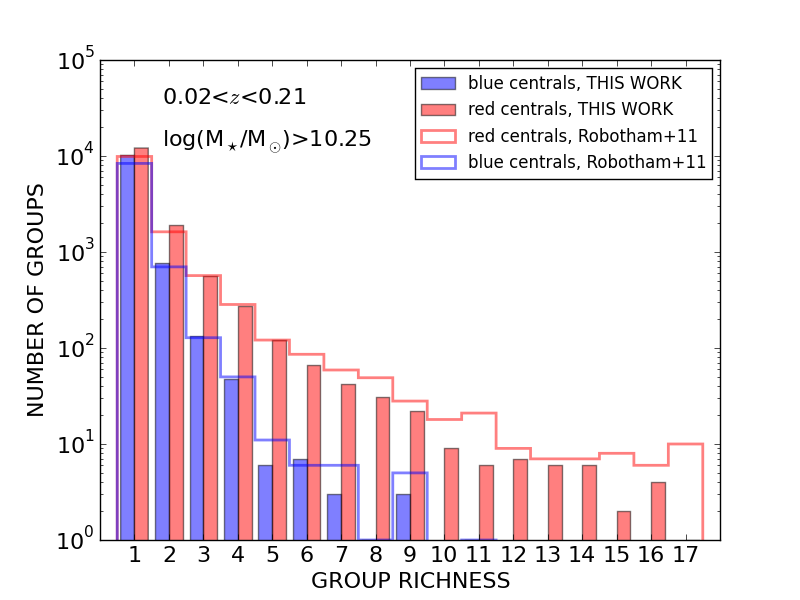}
\includegraphics[width=8.7cm]{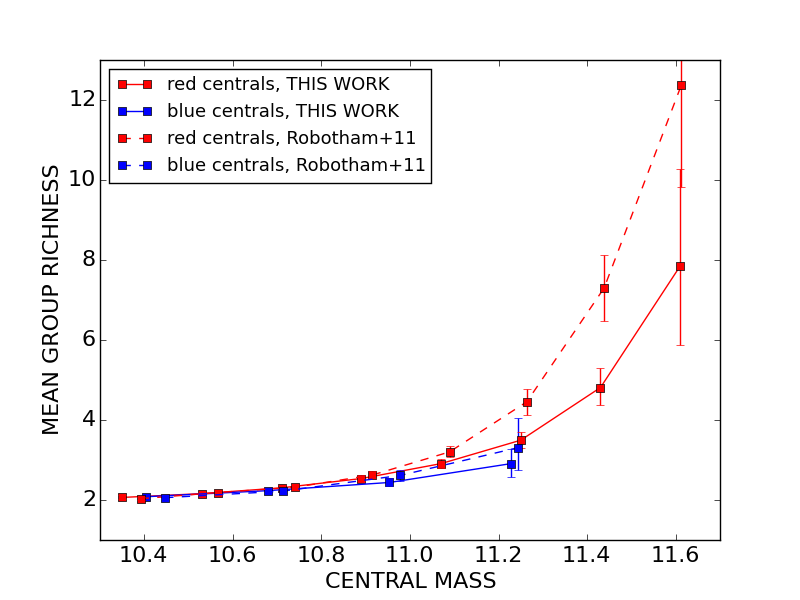}
\caption{{\bf Left:} The richness distribution of the groups derived in this work (Section \ref{sec:groups}) and by \citet{Robotham2011}. {\bf Right:} The mean group richness as a function of the central's stellar mass for groups with at least 2 members above the stellar mass limit. 76\% of group centrals are red in both group catalogs and collectively own $\sim 82\%$ ($85\%$ for \citet{Robotham2011}) of the satellites above the mass limit. }
\label{nfof_hist}
\end{figure*}

\begin{table}
\centering
\caption{Group catalog. \label{tab:galaxy_sample}}{No. of objects with $\Log (M_{\star} / M_\odot) > 10.25$ at $0.02<z<0.21$}
\begin{tabular}{cccccc}
\hline \hline
\multicolumn{2}{c}{isolated centrals} &  \multicolumn{2}{c}{group centrals} & \multicolumn{2}{c}{    satellites  } \\
              red & blue & red & blue & red & blue \\
\hline \hline
9659 & 8392 & 3092 & 975 & 4850 & 2378 \\
32.9\% &  28.6\% &  10.5\% &   3.3\% &  16.5\% &   8.1\% \\
\hline
\end{tabular}
\end{table}

\subsection{Central vs satellite classification}

Central galaxies are often assumed to be the most massive galaxies in a halo lying at the minimum of its gravitational potential well, while satellites are all remaining group members orbiting the centrals within the group potential. This central-satellite dichotomy is easily applied in simulations but represents a non-trivial challenge in real data. Groups may be fragmented, over-merged, they can contain interlopers or miss actual members, and spurious groups may be generated. 

To minimize the impact of group membership misidentification, in addition to a given physical property of galaxies (stellar mass and/or luminosity), some information about their spatial \citep[e.g.,][]{Robotham2011,Knobel2012,Knobel2015} and velocity \citep{Carollo2013} distribution within the group may be included. Following \citet{Robotham2011} \citep[see also][for a different implementation]{Eke2004}, we tested an iterative approach. The first step of this method is to compute the centre of mass of the group (\com), then to proceed iteratively by computing the projected distance (as in Eq. \ref{eq:d_perp}) from the \com for each group member and by rejecting the most distant galaxy. This process stops when only two galaxies remain: the most massive of the two is identified as the central galaxy of the group and all other group members are classified as satellites. As this iterative center coincides with the most massive galaxy in 98\% of the groups, we chose the most massive galaxies as centrals. Thus all groups keep their central when applying a mass cut-off, or else disappear completely. 

For each group in the redshift range $0.02<z<0.21$ with $\Nfof$ members, we defined $\Nfof^{m_{cut}}$ as the number of group members with mass $\Log (M_\star/ M_\odot)>10.25$ (see Section \ref{subsec:mass_completeness}). In the following sections, ``group galaxies'' refer to all galaxies in groups with $\Nfof^{m_{cut}}\geq 2$. ``Isolated", ``lone" or ``field" galaxies refer to galaxies for which $\Nfof=1$ instead of $\Nfof^{m_{cut}}=1$ to exclude known group centrals with satellites detected below the mass limit, although this makes negligible difference in our results. Of course group centrals with faint satellites undetected in the survey will still be present in the lone category, especially at the lower mass limit and upper redshift limit of the sample. The total numbers of blue and red isolated centrals, group centrals and satellites are reported in Table~\ref{tab:galaxy_sample}. Figure \ref{nfof_hist} shows the group richness distribution (left panel) and the mean group richness as a function of the central's stellar mass (right panel) for groups with blue and red centrals.

For comparison, we also show these distributions for the group catalog of \citet{Robotham2011}. Our algorithm leaves more galaxies alone (61.5\% vs 51.8\%), yields more small groups and less satellites (24.6\% vs 34.5\%) than \citet{Robotham2011}, who generate larger groups. 20\% of their satellites are field galaxies and 6\% group centrals in our own catalog. However but we find that these differences in the rates of fragmentation/merging is not significant enough to alter our statistical results (see also Section \ref{sec:caveats}). In both cases, 76\% of group centrals are red and these red centrals tend to have more satellites than blue ones of the same stellar mass, in agreement with \citet{WangWhite2012}. Red centrals collectively own over 80\% of the satellites above the chosen mass limit.


\section{Quenching and conformity}
\label{sec:results}

As mentioned in the introduction, \citet{Weinmann2006} first coined the concept of ``galactic conformity'', referring to the observation derived from a SDSS DR2  galaxy group catalog, that the fraction of quiescent satellites around a quiescent central was significantly higher than around a star-forming central. They observed this phenomenon at all satellite luminosities and at all halo masses, computed via abundance matching from the total luminosity of the groups. We explore this effect in the GAMA group catalog described in the previous section, as a function of stellar mass, of various local and large-scale environmental parameters, and finally of halo mass. 

\begin{figure}
\includegraphics[width=9.cm]{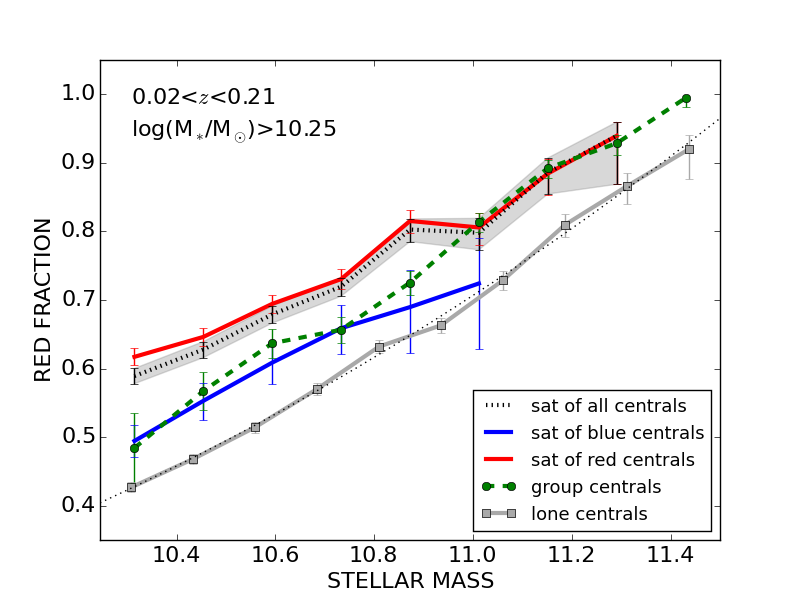}
\caption{The red fraction of galaxies as a function of stellar mass in units of $\Log (M_{\star} / M_\odot)$ in different environments as specified in the legend. The 1-$\sigma$ error bars are computed using the beta distribution quantile technique \citep{Cameron2011}. The excess quenching in satellites as well as in group centrals with respect to lone centrals suggests a form of  ``group quenching" is at play. The excess quenching in satellites of red centrals with respect to satellites of blue centrals is the galactic conformity signal. 
 }
\label{redfraction_mass}
\end{figure}

\begin{figure}
\includegraphics[width=9.cm]{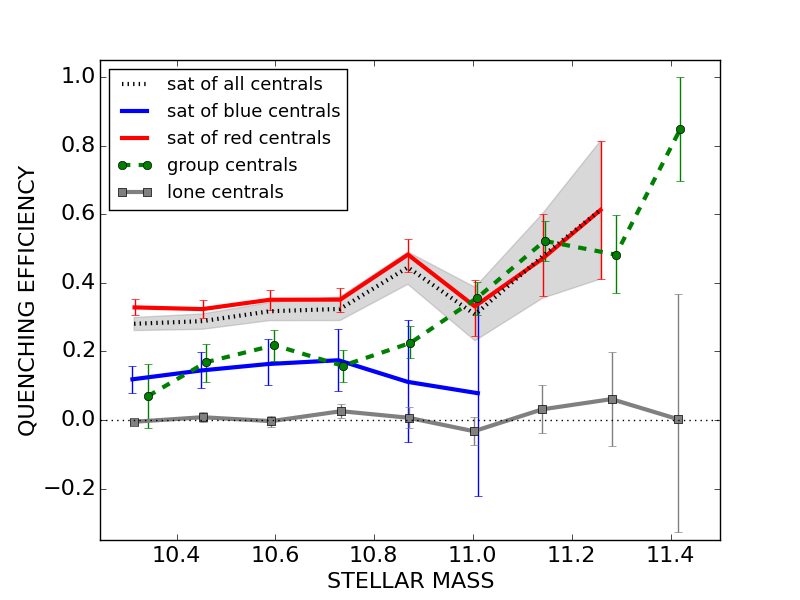}
\caption{The quenching efficiency (QE) of galaxies as a function of stellar mass in units of $\Log (M_{\star} / M_\odot)$ in different environments as in Fig.~\ref{redfraction_mass}. The 1-$\sigma$ error bars are based on 1000 bootstrap samples. This QE is defined as the excess red fraction of each population with respect to the red fraction of lone centrals, normalized by the star-forming fraction of that reference sample, at fixed stellar mass (Eq.~\ref{eq:eps_q_i}).  
It is, by definition, null for lone centrals. Up to $\Log(M_\star/M_\odot) \approx 11$, all the QE curves are nearly independent of mass. 
}
\label{qe_mass}
\end{figure}

\subsection{Mass quenching}
\label{subsec:mquenching}

We estimate the fraction of quiescent galaxies (red fraction) in a given sample as:
\begin{equation}
f_{q} =  \frac{ \sum_i  q_i w_i}    {\sum_i w_i },
\end{equation}  
where $q_i$ is unity if the $i^{\rm th}$ galaxy in the sample is quenched and zero otherwise. The parameter $w_i$ is the 1/$V_{\rm max}$ weight of the $i^{\rm th}$ galaxy (note that no weighting yields undistinguishable results above the chosen completeness limit). 

Figure \ref{redfraction_mass} shows the red fraction of galaxies as a function of stellar mass in different environments as follows: (i) isolated galaxies (gray solid line for the data and dotted line for its polynomial fit), (ii) centrals of groups (dashed green line), (iii) satellites of blue central galaxies (blue line), (iv) satellites of red central galaxies (red line) and (v) satellites of any central (dotted line with gray shaded errors). The fractions are estimated globally in each mass bin, not averaged over the red fractions of individual groups. In all environments, the red fraction increases strongly with increasing stellar mass. In the mass range that we are able to probe, a tenfold increase in stellar mass roughly doubles the red fraction. We will refer to this trend as ``mass quenching" \citep[e.g.][]{Peng2012}, as opposed to the ``environmental quenching" observed in the vertical direction at a given stellar mass.

\subsection{Conformity}
\label{subsec:conformity}

Figure \ref{redfraction_mass} shows that the red fraction of satellite galaxies, which is dominated by satellites of red centrals, is significantly higher at all masses than that of lone centrals. The difference is largest at low masses and getting smaller towards the highest masses, in general agreement with \citet{Bosch2008}, although the trend they observe is more dramatic (but see Section \ref{subsec:density}). 


The red fraction of satellites around blue centrals is about 10\% higher than the red fraction of isolated galaxies, except perhaps in the two highest mass bins ($\Log(M_\star/M_\odot) \gtrsim 10.8$) where uncertainties are large. At $\Log(M_\star/M_\odot) \gtrsim 11$, blue centrals become rare ($<20$\%) and we find no massive satellites around them. There are mixed results in the literature about whether or not satellites around star-forming central galaxies are quenched in excess of field galaxies. By examining the star formation properties of bright satellites around isolated Milky Way-like hosts in the local Universe, \citet{Phillips2014} found that quenching occurs only for satellites of quenched hosts while star formation is unaffected in the satellites of star-forming hosts. \citet{Hartley2015} also reported that the satellite population of star-forming centrals was similar to the field population of equal stellar mass at intermediate to high redshift. Conversely, \citet{Kawin2016} found a higher quenching efficiency of satellites around star-forming centrals compared to the background galaxies at $0.3< z <1.6$, in agreement with the study of \citet{Phillips2015} in the local Universe considering Milky Way-like systems hosting two satellites. Our analysis is more consistent with the latter studies, at least for low mass satellites.

Satellites of red centrals exhibit the highest level of quenching. Their red fraction is systematically about 10\% higher than that of satellites around star-forming centrals at all masses $\Log(M_\star/M_\odot) \lesssim 11$. This difference between the blue and the red lines is a significant galactic conformity signal, an environmental effect that adds to mass quenching: satellites around blue centrals have to be $\sim$ 2 times more massive than those around red centrals to exhibit the same level of quenching. In the high mass regime, $\Log(M_\star/M_\odot) \gtrsim 11$, where conformity disappears for lack of massive blue centrals hosting massive satellites, the red fraction becomes closer to that of field galaxies.

\subsection{Group quenching}
\label{subsec:gquenching}

Lastly, group centrals appear to follow the minimum quenching behavior of the satellite population as a function of mass: at $\Log(M_\star/M_\odot) \lesssim 11$, they quench like satellites of blue centrals, while at higher mass they converge with satellites of red centrals, in such a way that their red fraction runs $\sim 10 \%$  above that of isolated galaxies at all masses, reaching 100\% at  $\Log(M_\star/M_\odot) \gtrsim 11.4$. This excess quenching of group centrals over isolated centrals, despite the fact that the latter are likely to be contaminated by group members, supports the idea that quenching in groups is not reduced to ``satellite quenching'', as advocated by \citet{Knobel2015}. We refer to this mass-independent excess as ``group quenching", which, at this point, does not necessarily mean that satellites and group centrals respond equivalently to the group environment \citep{Knobel2015}. We return to this point later. 

\begin{figure} 
\hskip -0.4cm
\includegraphics[width=9.5cm]{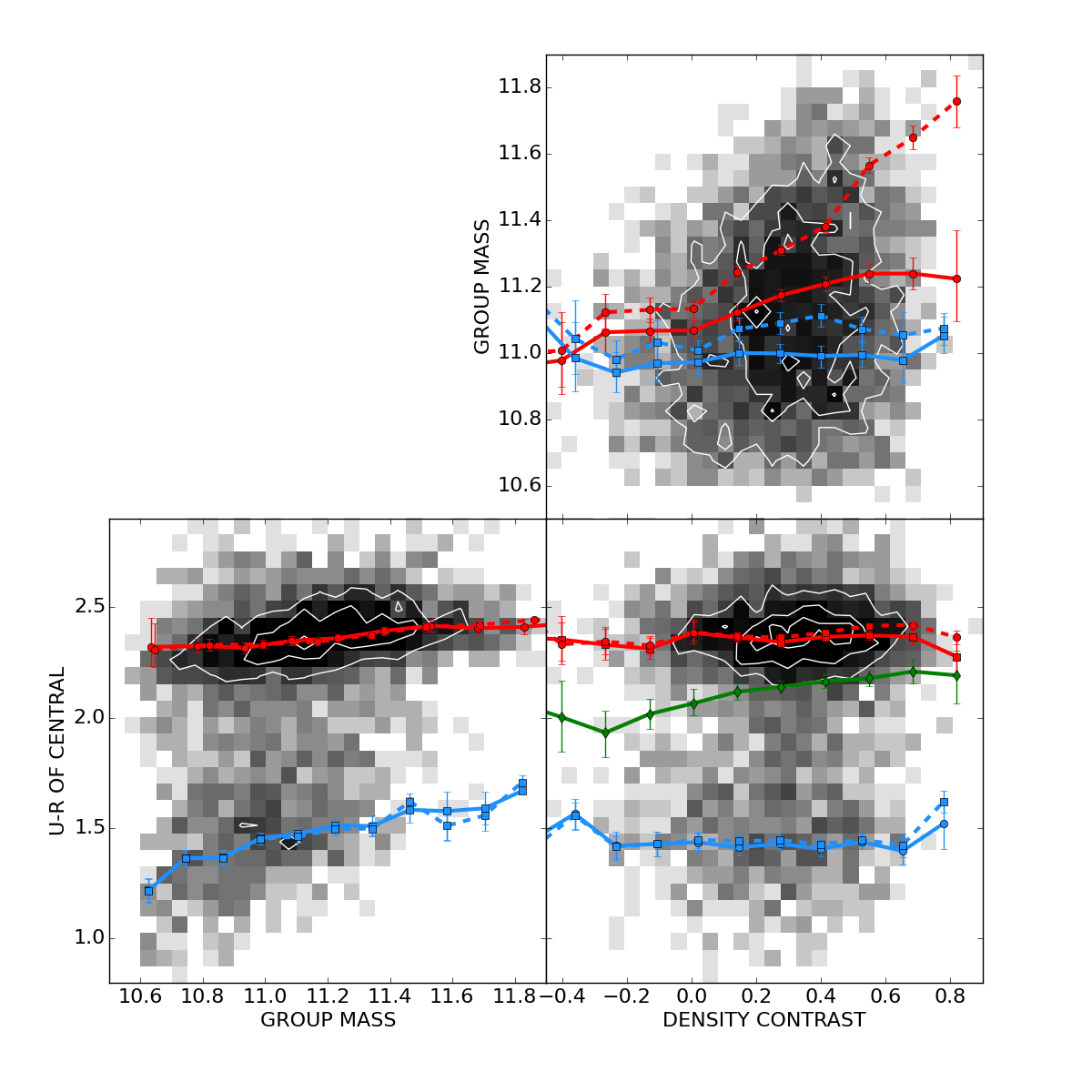}
\caption{Correlations between the three environmental parameters: dust corrected $(u-r)$ color of the central galaxies, group stellar mass and large scale density contrast at the central's location (see text for details). The solid blue and red lines show the mean values for blue and red group central galaxies respectively. The dashed blue and red lines are the mean values for satellites of blue and red centrals respectively (i.e. the same distributions weighted with the number of satellites in each bin). The green line shows the mean color/density relation of group centrals when blue and red ones are mixed.}
\label{environments}
\end{figure}

\begin{figure*}
\includegraphics[width=\textwidth]{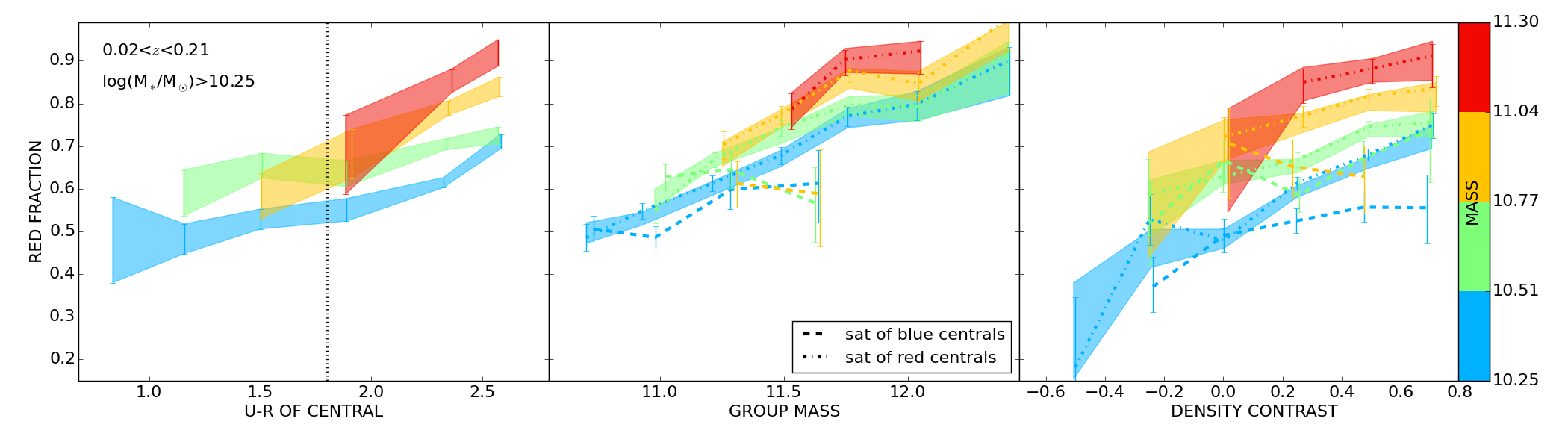}
\caption{The red fraction of satellites in four bins of stellar mass as a function of their central's $(u-r)$ color (left), their group's stellar mass in units of $\Log (M_{\star} / M_\odot)$ (middle), and the large scale density contrast in units of $\Log (1+\delta)$ (right). Mass quenching shows in the vertical direction at fixed environmental parameter. The upward trends with increasing central color, group mass and density, at fixed stellar mass, are essentially driven by satellites of red centrals (dot-dashed lines). The red fractions for satellites of blue centrals (dashed lines) are lower and consistent with being independent of all three environmental parameters. 
}
\label{redfraction_ur}
\end{figure*}

\begin{figure*}
\includegraphics[width=\textwidth]{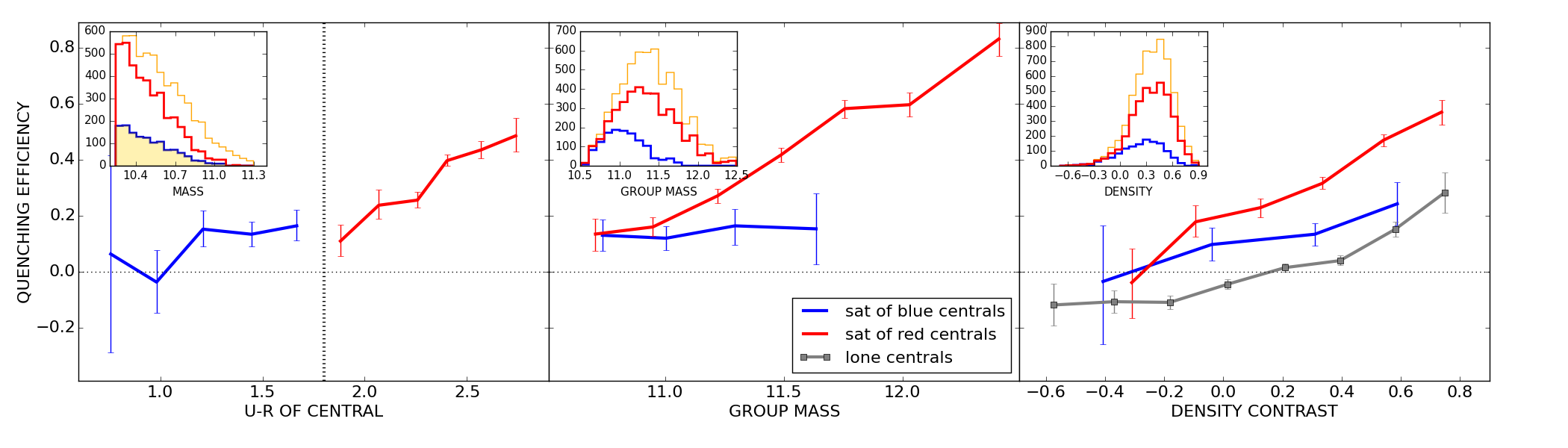}
\caption{Quenching efficiencies with respect to lone centrals at fixed mass (Eq.~\ref{eq:eps_q_i}) as a function of central color (left), group stellar mass (middle), and density contrast (right). The blue and red curves are the QE of satellites of blue and red centrals respectively in mass matched samples. The adopted shape for the mass distribution is shown in the inset of the left panel (shaded histogram), with the matched distributions for the blue and red central samples (unnormalized). The orange histogram is the parent stellar mass distribution of the satellites of red centrals. This QE formalism reinforces the conclusions of Fig.~\ref{redfraction_ur} with better statistics. The QE of lone centrals (gray curve in the right panel), which is calibrated over their own red fraction as a function of mass but independently of density, is also a strongly increasing function of density.  
}
\label{qe_ur}
\end{figure*}

Figure \ref{qe_mass} shows another view of these results, called quenching efficiency (QE or $\epsilon_{q}$), following the formalism of  \citep[]{Knobel2015} \citep[see also ][]{Bosch2008}. This formalism which separates out the dependence on stellar mass, helps to highlight the dependence on other parameters. The QE represents the excess red fraction of a given population with respect to the red fraction of a reference sample, $f_{q,\rm ref}$, normalized by the star-forming fraction of that reference sample: 

\begin{equation}
\label{eq:eps_q}
\epsilon_{q} (M_\star) =  \frac{f_{q}(M_\star) - f_{q,\rm ref}(M_\star)}    {1 - f_{q,\rm ref}(M_\star) }.
\end{equation}  
We choose the lone central population as our reference sample\footnote{This QE is interpreted as the probability that a central becomes quenched upon falling into another DM halo and becoming a satellite \citep{Knobel2013,Knobel2015}}, 
and fit its red fraction curve, $f_{q,\rm ref}(M_\star)$ (solid gray line in the left panel of Fig.~\ref{redfraction_mass}), with a polynomial function of order 2, $f_{q,\rm ref}^{\rm fit}(M_\star)$ (the overlapping dotted line):

\begin{equation}
f_{q,\rm ref}^{\rm fit}(M_\star) =  6.580-1.535\lgm +  0.091\lgmsq.
\end{equation}
This allows us to define the quenching efficiency of any individual galaxy $i$ of mass $M_\star$ as:

\begin{equation}
\label{eq:eps_q_i}
\epsilon_{q,i} (M_\star) =  \frac{q_i - f_{q,\rm ref}^{\rm fit}(M_{\star})}    {1 - f_{q,\rm ref}^{\rm fit}(M_{\star}) }.
\end{equation}

The meaning of these individual $\epsilon_{q,i}$ is limited ($\epsilon_{q,i}(M_\star)=1$ for red galaxies ($q_i=1$) and negative for blue galaxies ($q_i=0$)), but 
allows the QE of any set of galaxies to be computed as a function of any parameter of interest by simply averaging them over the sample \citep{Knobel2015}. 

Figure \ref{qe_mass} shows that this QE as a function of stellar mass, is, by definition, null for lone centrals. Up to the transition mass $\Log(M_\star/M_\odot) \approx 11$, all the QE curves are nearly independent of mass, with satellites of blue centrals and centrals of groups at the level of what we called group quenching (QE $\sim 0.15$). Galactic conformity manifests in the difference between the blue and the red curves, the QE of satellites around red centrals being about twice the group quenching value. In the higher mass regime where satellites of blue centrals no longer exist, the QEs of satellites of red centrals and of centrals of groups fast increase towards complete quenching (which can be inferred from the parallel behavior of their red fraction curves with respect to isolated galaxies in the left panel).

\section{Environmental effects}
\label{sec:environment}

\subsection{Environmental parameters}
\label{subsec:parameters}

We now address the dependence of the satellite red fractions and quenching efficiencies on three environmental parameters: 
\begin{itemize}
\item{the dust corrected $(u-r)$ color of central galaxies,}
\item{the group stellar mass $M_{gr}$, defined as the total stellar mass from group members with $\Log(M_\star/M_\odot)>10.25$,} 
\item{the density contrast ($1+\delta$) at the centrals' location, defined as the density of central galaxies -- satellites are excluded -- smoothed by a 3D Gaussian kernel of $\sigma=5$ Mpc and normalized by the redshift-dependent mean density of the survey.}
\end{itemize} 
These parameters are labelled ``U-R OF CENTRAL", ``GROUP MASS" and ``DENSITY CONTRAST" for visual clarity in all figures. The group mass and the density contrast are always expressed logarithmically as $\Log(M_{gr}/M_\odot)$ and $\Log(1+\delta)$, respectively. 

The density estimator used in satellite quenching studies is generally based on the ``fifth nearest neighbor" \citep[e.g.][]{Knobel2015}. 
A major disadvantage of this approach is that it doesn't probe the same environment for small and large groups \citep{Peng2012,Carollo2013}:  for small groups ($\Nfof \la 5$), the density is estimated on a scale much larger than the size of their DM halo, whereas for rich groups, it is measured well within their virial radius. Our choice of density estimator intentionally probes scales beyond the virial radius of all groups. 

Figure \ref{environments} shows how these environmental parameters may be correlated. The solid blue and red lines show the median values for blue and red centrals of groups respectively. The dashed blue and red lines are the median values for satellites of blue and red central galaxies respectively (i.e. the same distributions weighted with the number of satellites in each bin). A weak correlation is seen between group mass and color, expected since the group mass is dominated by the central's mass, which correlates with color (Fig.~\ref{cmd}). None is found between density and color. A color-density relation appears only when star-forming and quenched centrals are mixed (green line), reflecting their evolving ratio with density (see Section \ref{subsec:density}).
Density does correlate with group mass for groups with red centrals. The correlation is all the more apparent when the medians are weighted by the number of satellites in each group (solid versus dashed lines): rich, massive groups with red centrals are clearly more common in high density environments. The satellite weighting effect needs to be kept in mind when considering environmental correlations in general. Artificially boosted trends may be induced by a few very rich groups, as emphasized by \citet{Sin2017} in the case of 2-halo conformity signals.

\subsection{Satellite quenching}
\label{subsec:satquenching}

Figure \ref{redfraction_ur} shows the red fraction of satellite galaxies as a function of the above three environmental parameters in four bins of stellar mass, which displays mass quenching in the vertical direction. The left panel shows the red fraction of satellites as a function of their central's color, with the vertical dotted line indicating our boundary between blue and red centrals. This expands the blue/red central dichotomy in Fig.~\ref{redfraction_mass} to a continuum in central color, similar to the trend in \ssfr reported by \citet{Knobel2015}. The fraction of red satellites increases in all mass bins, most noticeably in the lowest, as their centrals redden, in such a way that the red fraction of low mass satellites around very red centrals exceeds that of satellites several times more massive around very blue centrals. 

More significant trends are found with the groups' stellar mass (middle panel), in agreement with earlier studies showing that satellite quenching is more efficient in more massive systems \citep[using the mass or the magnitude of the central galaxy or some proxy for the halo mass;][]{Weinmann2006,Ann2008,Prescott2011,Knobel2015}. 
In addition, two distinct sequences for satellites of blue and red centrals appear (dashed and dot-dashed lines respectively): satellites of similar stellar mass, in groups of similar group mass above $\sim 10^{11}M_\odot$, are more likely to be red if their central is red than if it is blue, as also found by \citet{Knobel2015} using the central's mass and halo mass. The upward trends with group mass are essentially driven by satellites of red centrals, which vastly dominate the satellite population. The curves for satellites of blue centrals, which span a narrower range of group masses and have poorer statistics, are consistent with being flat. 

Lastly, significant correlations are observed between the red fractions in all mass bins and the density contrast (right panel). Again two sequences are seen that exhibit conformity at fixed density contrast, at least for $\Log(1+\delta)>0$. For satellites of red centrals, quenching increases significantly with $\Log(1+\delta)$, while the statistics are inconclusive for satellites of blue centrals. These trends may simply duplicate the middle panel: satellite quenching is enhanced in massive groups whose central is red and these are found preferentially in high density regions (cf. Fig.~\ref{environments}). We will attempt to disentangle these effects in the next section. 

\begin{figure*}
\includegraphics[width=\textwidth]{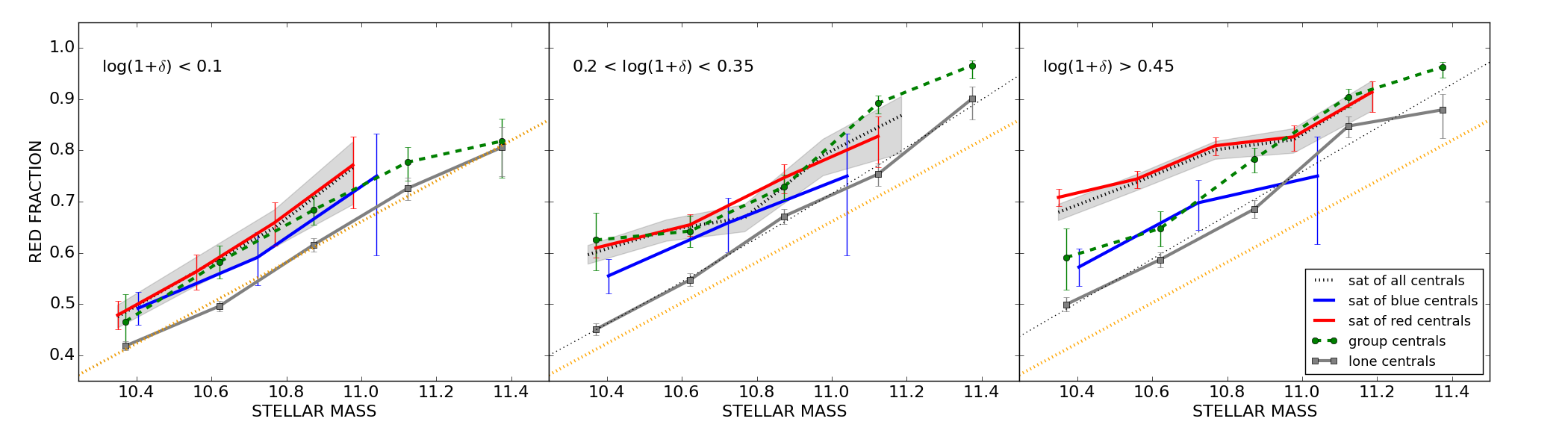}
\caption{Same as the left panel of  Fig.~\ref{redfraction_mass} in 3 bins of density contrasts:  $\Log(1+\delta)<0.1$ (left), $0.2<\Log(1+\delta)<0.35$ (middle) and $\rm log(1+\delta)>0.45$ (right). The dotted orange line in all 3 panels fits the lone central curve in the lowest bin to guide the eye. The vertical rise as density increases is clear for all galaxies at all masses, indicating that quenching is affected by the environment far beyond the virial radius of DM halos. At low and medium density, conformity is marginal and group centrals do not distinguish themselves from satellites, supporting the idea of  group quenching by \citet{Knobel2015}. Conformity is most significant in the highest density bin.
}
\label{redfraction_mass_dens}
\end{figure*}

\begin{figure*}
\includegraphics[width=\textwidth]{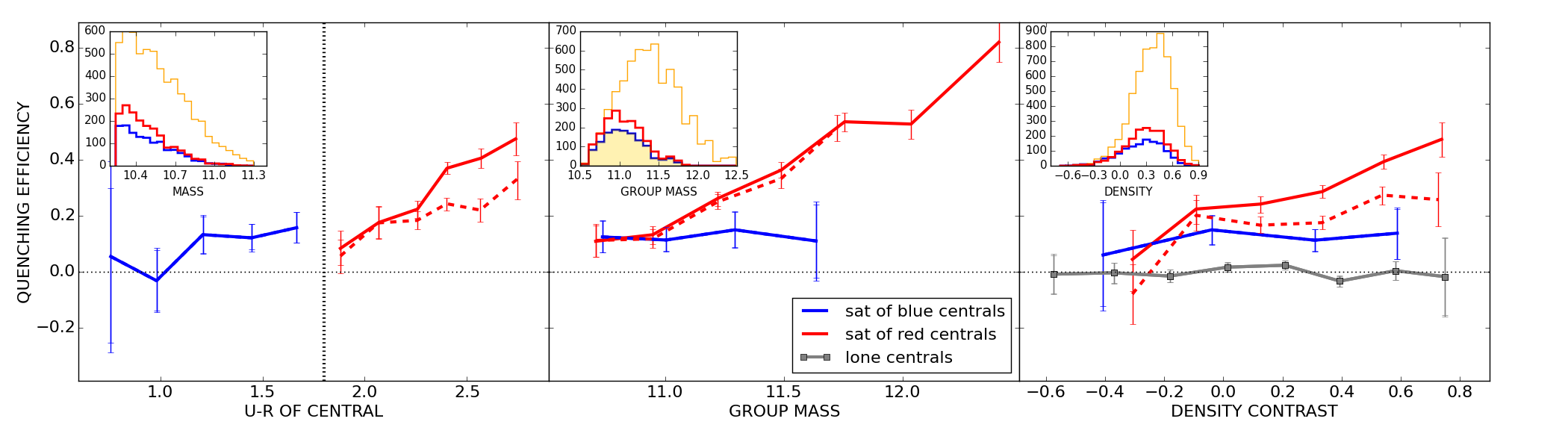}
\caption{Quenching efficiencies with respect to lone centrals at fixed mass {\it and} density (Eq.~\ref{eq:eps2_q_i}) as a function of the three environmental parameters. In solid lines, the satellites of red centrals are picked to have the same stellar mass distribution as the satellite of blue centrals; in dashed lines, they have the same distribution of group stellar mass (yellow shaded, blue and red histograms in insets). The orange histograms are the parent distributions of the satellites of red centrals. The solid gray curves show the new QE of lone centrals, designed to be null with respect to both stellar mass and density. Most of the conformity signal originates from comparing groups of different stellar masses. 
}
\label{qe_ur_dens}
\end{figure*}

\begin{figure}
\includegraphics[width=9cm]{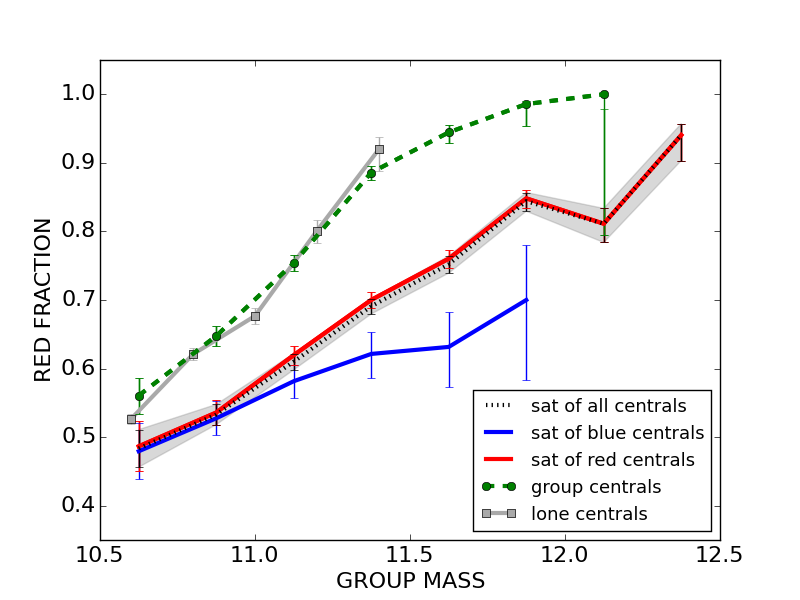}
\caption{The red fraction of galaxies as a function of group stellar mass, which, for field galaxies, equals their stellar mass by definition. The red fraction gap between group centrals and lone centrals at fixed stellar mass is filled at fixed group mass, as if group centrals ``carried" the extra weight of their satellites. Conformity is present at fixed group mass as expected from Fig.~\ref{qe_ur_dens}.
}
\label{redfraction_groupmass}
\end{figure}

Since satellites behave similarly in the four mass bins we probed, we may increase our statistics, especially for satellites of blue centrals, by using mass-matched samples instead of bins, i.e. samples of satellites of blue centrals and of red centrals having similar stellar mass distributions. We also make use of the QE formalism described in the previous section, to detach mass quenching from the environmental quenching we are trying to highlight. Figure \ref{qe_ur} recasts Fig.~\ref{redfraction_ur} using this new methodology. The adopted common mass distribution is the intersection of the two samples, i.e. the mass distribution of satellites of blue centrals. The unnormalized blue and red distributions are shown in the leftmost inset, with the intersection shaded in yellow. The orange histogram is the total mass distribution of the satellites of red centrals. As it is significantly larger than the blue histogram, we make several draws in it in order to pick every galaxy at least once, and compute the final QE curve as the average of the QE curves computed for each individual draw. 

The three panels of Fig.~\ref{qe_ur} reinforce the findings of Fig.~\ref{redfraction_ur}: the QE of satellites is a smoothly rising function of central color from blue to red; the QE of satellites around red centrals increases with their group stellar mass, while it is independent of it for satellites of blue centrals, but also consistently positive (i.e. they quench more efficiently than field galaxies); the QE of both populations strongly increases with large scale density. This figure also emphasizes that much of the excess quenching in satellites of red centrals originates from the most massive groups that have no counterpart with blue centrals. The QE of lone centrals - which is calibrated over their own red fraction as a function of mass but independently of density - is also a strongly increasing function of density contrast, negative below the peak density, positive above. This point is addressed in the next section. We note too that there exists circumstances in which conformity disappears, i.e. the blue and red curves converge: first for centrals in the ``green valley", the QE being a monotonically increasing function of central color from the bluest to the reddest;  secondly in low mass groups  ($M_{gr} \lesssim 10^{11}M_\odot$) and thirdly at the lowest density contrasts ($\Log(1+\delta) \lesssim 0$) where the QE of both populations reaches zero or negative values. The last two circumstances are clearly correlated (Fig.~\ref{environments}). We will now attempt to disentangle them.

\subsection{Density quenching}
\label{subsec:density}


Figure \ref{redfraction_mass_dens} reproduces the left panel of Fig.~\ref{redfraction_mass} in three bins of density contrast.  The middle bin spans the peak of the density distribution of the full sample.
The dotted orange baseline in the three panels is a fit to the lone central curve in the lowest bin to guide the eye: the uplift as density increases is clear for all galaxies, indicating that quenching, is affected by the environment far beyond the virial radius of DM halos\footnote{Qualitatively similar results are also found using densities computed on a scale of 8 Mpc instead of 5.}. 

For central galaxies, this effect is interpreted as reflecting the earlier collapse of proto-halos in large-scale overdense regions. At a given halo mass, the halos populating denser environments are older on average, with different accretion histories (delayed or quenched mass inflow), a phenomenon referred to as assembly bias \citep{Sheth2004,Croton2007}. The age of a halo is usually defined as the epoch at which the halo has assembled one half of its current mass, however \citet{Tinker2017} showed that the amplitude of assembly bias was significantly reduced if age was defined using halo mass at its peak rather than current value (removing the effect of ``splashback" halos), and in better agreement with their analysis of SDSS data. They find a $\sim 5\%$ increase  from low to high large-scale (10 $h^{-1}$Mpc) density in the red fraction of central galaxies with $\Log(M_\star/M_\odot) \gtrsim 10.3$. The increase between the orange and gray dotted line in the right panel of Fig.~\ref{redfraction_mass_dens} is $\sim 10\%$, a reasonable agreement given the many differences in the two analyses. For comparison, a factor of 2 in stellar mass at fixed density roughly induces the same increase in red fraction.

Also notable in this figure is the increasing vertical gradient between the satellites of red centrals and the other curves, in particular that of satellites of blue centrals, at  $\Log(M_\star/ M_\odot) \la 11$ as density increases: conformity is hardly detectable in the lowest density bin, emergent in the middle bin, and very significant in the highest bin, where it is also mass dependent. The difference between satellites of red centrals and lone centrals in this bin is  significantly larger at lower masses than in the average density case (Fig.~\ref{redfraction_mass}), in better agreement with \citet{Bosch2008} (their Fig.~8, top right panel, for all satellites combined).

In all three panels, the red fraction curve of group centrals runs roughly $10\%$ higher than that of lone centrals at all masses, as in the global case of Fig.~\ref{redfraction_mass}. Thus group quenching, which we defined in Section \ref{subsec:gquenching} as the difference between the dashed green curve and the gray curve, adds a somewhat constant boost to density quenching. At low and medium density, where conformity is marginal, group centrals do not distinguish themselves from the satellite population as a whole, in agreement with \citet{Knobel2015} and in support of their group quenching definition, whereby, in the restricted part of the mass and (5$^{th}$ nearest neighbor) density parameter space that they share in the SDSS, group centrals and satellites ``feel environment in the same way". However at high density, satellite quenching, which is dominated by satellites of red centrals, far exceeds the red fraction of group centrals at $\Log(M_\star/M_\odot)\lesssim11$. 

To separate out the effect of large-scale density, we fit the red fraction of lone galaxies as a function of both mass and density in the following empirical way: 
\begin{equation}
\label{eq:fq_md}
 \tilde f_{q,\rm ref}(M_\star,\delta) =  f_{q,\rm ref}^{\rm fit}(M_\star) \times (1+g(\delta)). 
 \end{equation}
In practise, $f_{q,\rm ref}^{\rm fit}(M_\star)$ is a new polynomial fit (of order 2) to the red fraction of lone galaxies in the middle bin (the gray line in the middle panel of Fig.~\ref{redfraction_mass_dens}):
\begin{equation}
f_{q,\rm ref}^{\rm fit}(M_\star)  =  5.43 -1.31\lgm+ 0.08\lgmsq,
\end{equation}
and $g(\delta)$ is a polynomial fit (of order 3) to the QE curve of lone galaxies as a function of $\Delta=\Log(1+\delta)$ (the equivalent of the gray line in Fig.~\ref{qe_ur} using the new $f_{q,\rm ref}^{\rm fit}(M_\star)$), multiplied by a fudge factor of 0.8:
\begin{eqnarray} 
g(\delta)   =  0.8\times( -0.07+0.23\Delta+ 0.25\Delta^2 -0.02\Delta^3)
\end{eqnarray}
This empirical recipe provides a good fit to the red fraction of lone galaxies as a function of both mass and density. It allows us to redefine the quenching efficiency $\epsilon_{q,i} (M_\star,\delta)$ of an individual galaxy $i$ of mass $M_\star$ living in a region of density contrast $\delta$, as:
\begin{equation}
\label{eq:eps2_q_i}
\epsilon_{q,i} (M_\star,\delta) =  \frac{q_i - \tilde f_{q,\rm ref}(M_{\star},\delta)}    {1 - \tilde f_{q,\rm ref}(M_{\star},\delta) },
\end{equation}
and the QE of a galaxy sample as the average of these individual $\epsilon_{q,i}$.

The solid lines in Fig.~\ref{qe_ur_dens} represent this new QE for satellites of blue and red centrals as a function of the three environmental parameters in stellar mass-matched samples as in Fig.~\ref{qe_ur}. The gray curve in the right panel represents the QE of lone centrals, designed to be null with respect to both stellar mass and density. The dependence on density completely disappears for satellites of blue centrals, whereas is remains significant for satellites of red centrals. 

As was already visible in Fig.~\ref{qe_ur}, much of the excess quenching in the satellites of red central population originates from the most massive groups that have no blue central counterpart. The dashed lines in Fig.~\ref{qe_ur_dens} show that if we match the group stellar mass distributions of satellites around blue and red centrals, which excludes these massive groups and which also results in nearly perfectly matching the stellar mass and density distributions of both populations (distributions shown in the insets), the dependence with density and color are strongly reduced for the remaining satellites of red centrals. Some amount of conformity persists as a function of group mass, which affects the QE of satellites of red centrals only, but this figure allows us to conclude that most of the conformity signal arises from the most massive groups with the reddest centrals that non only have no counterpart with blue centrals in terms of group mass, but in which satellite quenching is more dependent on density than in other groups, including low mass groups with red centrals.

In Fig.~\ref{redfraction_groupmass}, we show the red fraction of all galaxies as a function of group stellar mass. For field galaxies, the group stellar mass equals their stellar mass by definition. We find that the quenching gap between group centrals and lone centrals at fixed stellar mass is filled at fixed group mass, as if group centrals ``carried" the weight of their satellites (an extra $\sim 0.25$ dex in stellar mass on average). Satellites, although they are also boosted by the group environment, do not ``feel'' the added weight of their more massive central: at a given group mass, they remain significantly less quenched than their central, as expected from their having lower masses, with a conformity effect expected from the previous figure. Since the QE efficiency of satellites of blue centrals is independent of group mass, the positive slope of their red fraction in this figure may simply be attributed to their increasing mean stellar mass and surrounding density as group mass increases. We go back to this point in Section \ref{subsec:halo}. 



\begin{figure}
\includegraphics[width=8.5cm]{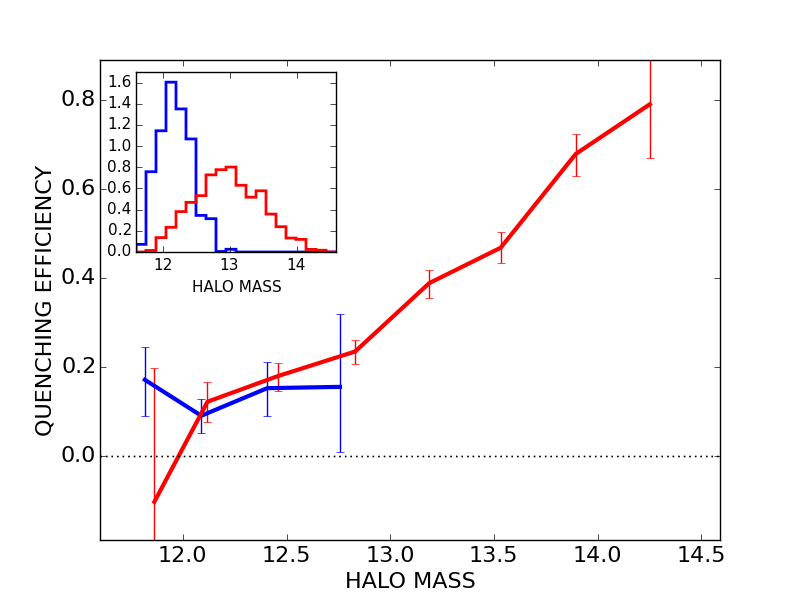}
\caption{ The QE of satellites around blue and red centrals (blue and red curves respectively), as a function of halo mass in units of $\Log(M_{h}/M_\odot)$ estimated from the blue/red halo-to-stellar-mass ratios of \citet{ZuMandelbaum2015}  (see text for details). The inset shows the normalized halo mass distributions for satellites of blue and red centrals. This model highlights two regimes: at $\Log(M_{h}/M_\odot) \gtrsim 13$, centrals are fully quenched and conformity has no meaning, while at $\Log(M_{h}/M_\odot) \lesssim 13$, central quenching is still ongoing and conformity is insignificant.
}
\label{shmr}
\end{figure}


\begin{figure*}
\includegraphics[width=\textwidth]{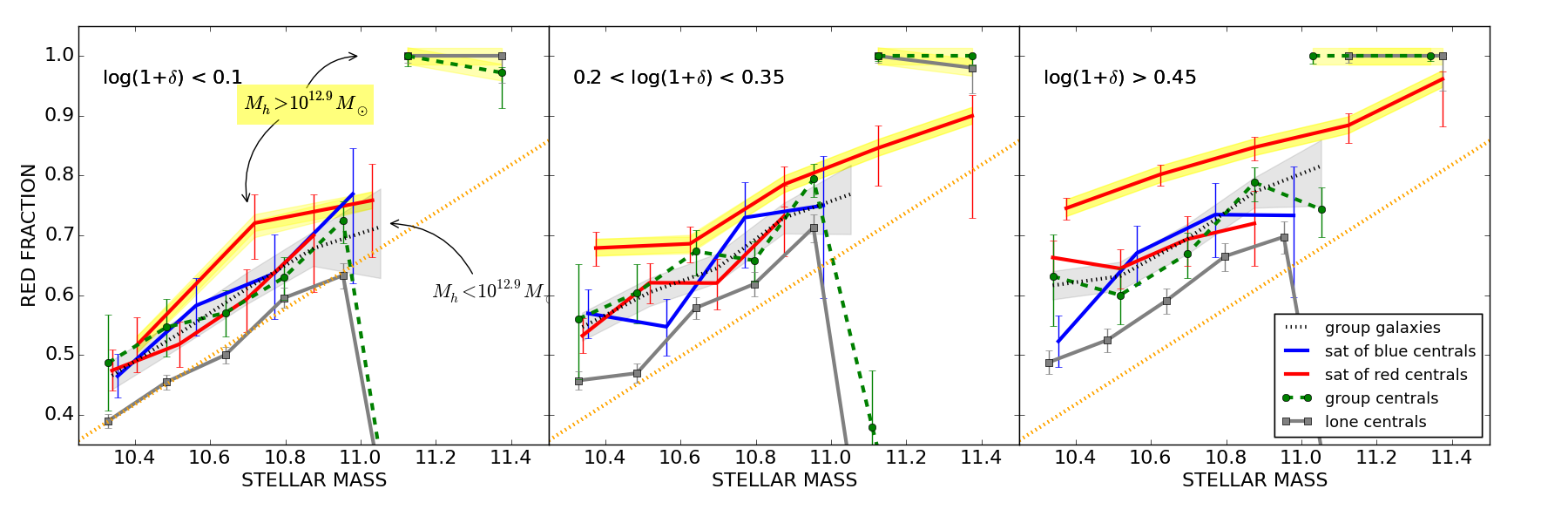}
\caption{The red fraction of satellites as a function of stellar mass in 3 bins of density as in Fig.~\ref{redfraction_mass_dens}, and in two regimes of halo mass: $\Log(M_{h}/M_\odot)<12.9$ and $>12.9$. In the low halo mass regime, conformity is insignificant and satellites and group centrals (collectively shaded in gray) exhibit the same red fraction excess over field galaxies at all mass $\Log(M_{\star}/M_\odot)\lesssim11$ and density, in agreement with \citet{Knobel2015}. In the high halo mass regime (highlighted curves), the centrals are massive, fully quenched galaxies, and their satellites undergo significantly more quenching than their counterparts in smaller halos, all the more so as density increases, giving rise to the increasing conformity signal observed in Fig.~\ref{redfraction_mass_dens}.
 }
\label{redfraction_halosplit}
\end{figure*}

\begin{figure}
\includegraphics[width=9cm]{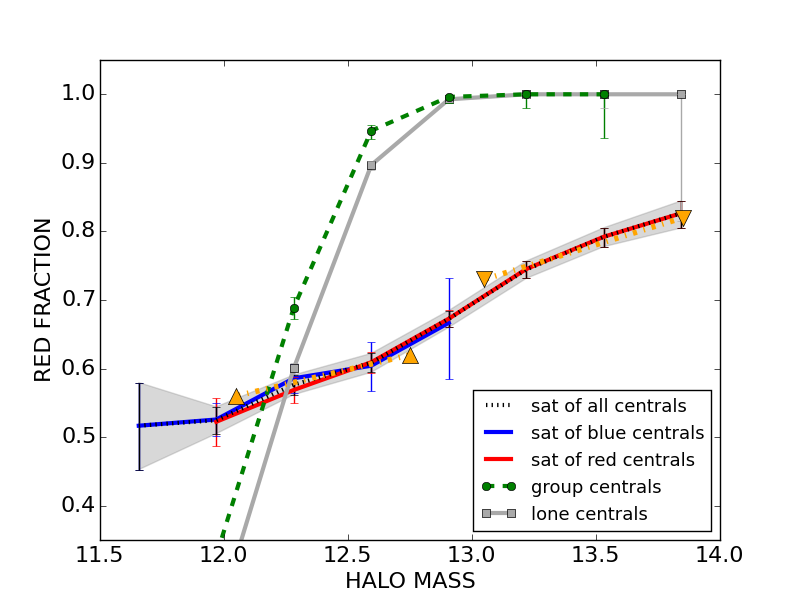}
\caption{The red fraction of centrals and satellites as a function of halo mass.  The orange triangles are the predictions based on the stellar mass and density dependence of the satellite red fraction in the two regimes of halo mass (see text for details).
}
\label{redfraction_halomass}
\end{figure}

\begin{figure}
\includegraphics[width=9cm]{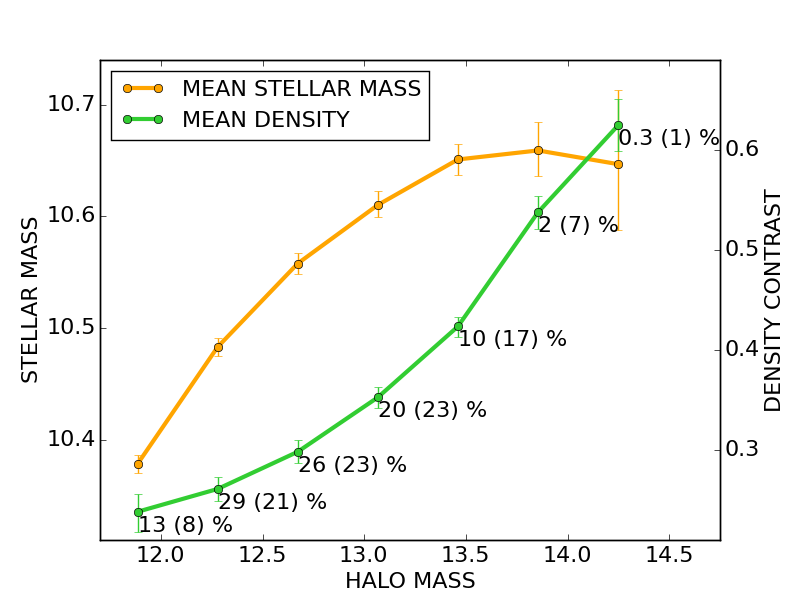}
\caption{Correlations between the halo mass and the mean surrounding density and the mean stellar mass of the satellites. The numbers are the fraction of groups and of satellites (in parenthesis), in each bin of halo mass. Most of the conformity signal in Fig.~\ref{redfraction_mass} originates from a small minority of groups with high halo mass at high density.
}
\label{halo_dens}
\end{figure}

\subsection{Halo quenching}
\label{subsec:halo}

Assuming that group mass is a good proxy for halo mass, our results confirm that galactic conformity is indeed observed at fixed halo mass as previously claimed \citep{Weinmann2006,Hartley2015,Knobel2015,Kawin2016}. A widely discussed explanation for this has been the ``assembly bias"  \citep{Gao2005,Croton2007}, which refers to the dependence of halo clustering on properties other than mass. One such property is age: older halos (halos that assembled earlier) presumably cluster more strongly than younger halos (that formed later) of the same mass, the effect being stronger for less massive halos. Another property, which correlates with halo age, is concentration: more concentrated halos are typically more clustered than less concentrated ones.

If there exists a sufficiently tight relation between one of these halo properties and galaxy color or star-formation history, galaxy conformity is expected to arise. A relation between galaxy color and halo age was indeed found in the cosmological hydrodynamical simulation Illustris \citep{Bray2016}, where the reddest galaxies preferentially reside in the oldest halos.  Similarly, \citet{Paranjape2015} and \citet{Pahwa2016} found conformity in models correlating galaxy color with halo concentration. 

However inferring halo masses in real data is an uncertain task and those derived from total group luminosities or masses, as used in establishing galactic conformity, are not flawless. In fact it was observed that red galaxies reside in more massive halos than average galaxies of similar luminosity or mass \citep{Zehavi2011,WangWhite2012,Krause2013}. More recently, \citet{ZuMandelbaum2016} claimed to have discovered, from weak lensing measurements of local bright galaxies in the SDSS, that there exists a strong bimodality in the average host halo mass of blue versus red galaxies: at fixed stellar mass, red centrals preferentially reside in halos that are a factor of 2 to 10 more massive than halos hosting blue galaxies. They found that a model in which ``halo quenching" -- referring to all the physical processes tied to halo mass (virial shocks, accretion shocks, AGN feedback, gas stripping) -- is the main driver of galaxy quenching, best fits the available data and explains galactic conformity without assembly bias \citep{WangWhite2012,Phillips2014, Phillips2015,ZuMandelbaum2017}. 



To test this scenario, we derive halo masses for all groups using the double, blue/red, average halo-to-stellar-mass-ratio (HSMR) of \citet{ZuMandelbaum2015} (their Fig.~16), applied to the central galaxies. Figure \ref{shmr} shows the quenching efficiency of satellites as a function of this halo mass estimate: conformity at fixed halo mass has disappeared. This model splits the halo mass into two distinct regimes that were blurred with the group stellar mass: a high halo mass regime, $\Log(M_{h}/M_\odot) \gtrsim 13$ (29\% of the groups, 46\% of the satellites, 5\% of the field galaxies), in which centrals are fully quenched and conformity has no meaning, and a low halo mass regime, $\Log(M_{h}/M_\odot) \lesssim 13$ (71\% of the groups, 54\% of the satellites, 95\% of the field galaxies), in which centrals still experience quenching, i.e. can be either red or blue at a given halo mass, but with no significant impact on the quenching of their satellites. The satellite QE in this regime is also consistent with being independent of halo mass. 

Figure \ref{redfraction_halosplit} revisits Fig.~\ref{redfraction_mass_dens}, in the two regimes of halo mass. We note that halos of both types exist in all tree density bins although they are not equally represented ($\sim 24,\ 25,\ 22\%$ of groups in the low mass regime are in the low, medium and high density bins respectively vs. $\sim$ 13, 28, 31\% for groups in the high halo mass regime). 
In the low halo mass regime, conformity is insignificant indeed. Satellites and group centrals also exhibit the same quenching excess ($\sim$ 10\%) over field galaxies at all mass and density, in agreement with the group quenching definition of \citet{Knobel2015}. In the high halo mass regime, in which the centrals are all massive ($\Log(M_{\star}/M_\odot)>11$), fully quenched galaxies, satellites undergo significantly more quenching than their counterparts in smaller halos. This effect increases with density, giving rise to the increasing conformity signal observed in Fig.~\ref{redfraction_mass_dens} from low to high density when the two regimes are mixed. The dependence on stellar mass also appears to be less significant than for galaxies in low halo mass groups and field galaxies (except in the leftmost bin of both stellar mass and density, but this part of the parameter space is poorly sampled). We stress that the two regimes, brought out by the two \citet{ZuMandelbaum2015} HSMRs, are equivalent to setting apart groups with massive red centrals (in the top right tip of Fig.~\ref{cmd}) from all the others {\it regardless of halo mass}. 

Figure \ref{redfraction_halomass}  shows the red fractions of both satellites and centrals as a function of halo mass. The red fraction of group and lone centrals (matched in stellar mass distribution) is a much steeper function of this halo mass than of the stellar group mass (Fig.~\ref{redfraction_groupmass}), reaching unity at $\sim 10^{13}M_\odot$, while all satellites follow the same gentler path regardless of central type. Since the red fraction of satellites and group centrals depends on their own stellar mass equivalently in low mass halos, and halo mass is strongly correlated to the stellar mass of the central only, it is not surprising that central quenching depends strongly on halo mass while satellite quenching does not, at least in the low halo mass regime. The slope might simply be explained by the increasing mean stellar mass and surrounding density of the satellites as halo mass increases. At $\Log(M_{h}/M_\odot) \sim 12$ and 12.7,  the mean stellar masses of the satellites are $<\Log(M_{\star}/M_\odot)>=$ 10.41 and 10.54 respectively, and their mean surrounding densities are 0.24 and 0.30. Using Eq. \ref{eq:fq_md}, which models the red fraction of the field population as a function of mass and density, to which we add 0.1 to account for the group quenching excess, we compute the mean red fractions expected from these two combinations of stellar mass and density. The resulting values are shown as orange, upward pointing triangles. The reasonable agreement indicates that satellite quenching depends on halo mass inasmuch as more massive halos contains more massive satellites on average (the difference in mean density makes negligible difference).  
At $\Log(M_{h}/M_\odot) \sim 13$ and 13.8, in the high halo mass regime, the mean stellar masses of the satellites are $<\Log(M_{\star}/M_\odot)>=$ 10.61 and 10.65 while their mean surrounding densities are 0.37 and 0.6, respectively. The saturation in mean stellar mass is expected from the increasingly small contribution of high mass satellites to the overall mass distribution. Here we simply compute the mean red fractions of satellites in small bins around these two combinations, with the condition that their central is red and massive. The resulting values are shown as orange downward triangles. The agreement shows that the dependence on halo mass in this regime is equivalent to a dependence on density. Figure \ref{halo_dens} shows how the two properties are strongly correlated, as in the case of the group stellar mass shown in Fig.~\ref{environments} for groups with red centrals.


We conclude that halo mass, as defined in this section, is the ``hidden" parameter \citep{Knobel2015} behind galactic conformity, due to comparing satellites in two distinct ranges of halo mass. The \citet{ZuMandelbaum2015} HSMRs separate groups with massive ($\Log(M_{\star}/M_\odot)>11$), red centrals from all the others, and offers a halo mass interpretation to our finding that satellite quenching appears to proceed differently in these two types of groups. The change of regimes marks a change in the stellar mass and density dependence of satellite quenching, from mostly stellar mass dependent to mostly density dependent. We find that the concept of group quenching advocated by \citet{Knobel2015}, whereby satellites and group centrals quench similarly at a given stellar mass and density (``feel environment in the same way''), holds in the low halo mass regime ($\sim 70\%$ of the groups): the red fraction of all galaxies in these groups is determined by the same combination of their own stellar mass and large scale density, with a constant $\sim 10\%$ boost with respect to their field counterpart. In more massive halos whose centrals are massive, fully quenched galaxies ($\sim 30\%$ of the groups), satellite quenching also depends on stellar mass and surrounding density, but less on the former and more on the latter, which strongly correlates with halo mass. This may point to different, satellite-specific quenching processes in these massive halos, or simply to the increasing importance of group pre-processing, i.e. satellites first quenching in lower mass groups prior to infall into more massive ones, as halo mass increases \citep{Wetzel2013}.

\begin{figure}
\includegraphics[width=9.cm]{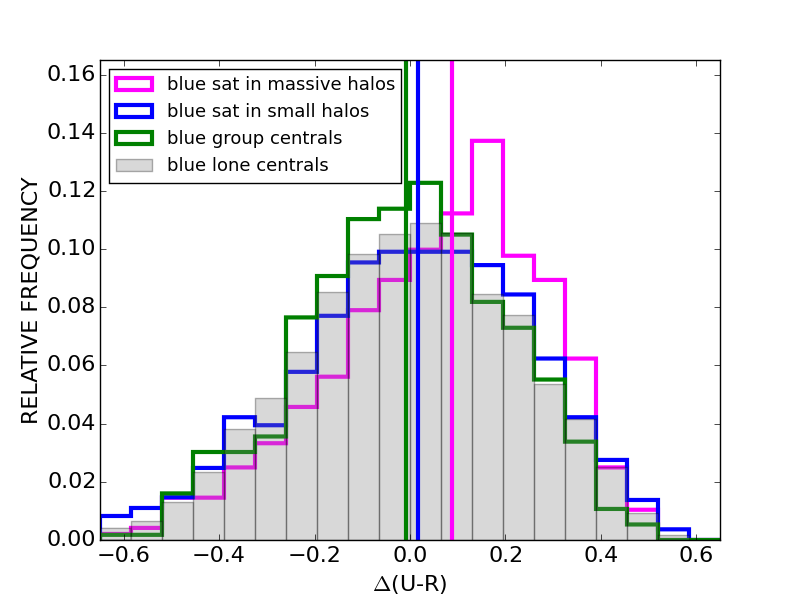}
\caption{The $\Delta(u-r)$ distributions (Eq.~\ref{delta_ur}) of blue field galaxies and blue satellites in low mass and high mass halos ($\Log(M_{h}/M_\odot) < $ or $> 12.9$) respectively, with corresponding median values (vertical lines). The star-forming activity of blue satellites in low mass halos ($\sim 70\%$) is similar to that of blue isolated galaxies, while it is mildly suppressed for blue satellites in massive halos ($\sim 30\%$). 
}
\label{grootes_halo}
\end{figure}

\section{Star-formation in groups}
\label{sec:sf}

We now examine the level of star-forming activity of blue satellites compared to their field counterparts. \citet{Grootes2017} found that the median \ssfr-stellar mass relation of a morphologically selected sample of disk-dominated satellites in the GAMA survey was mildly suppressed compared to isolated spiral galaxies. They also find that this mild suppression originates from a minority population ($\lesssim 30\%$) with strongly suppressed \ssfr with respect to isolated spirals, whereas the majority of spiral satellites show no sign of being affected by their group environment. Nor do spiral group centrals. These results led them to conclude that the gas cycle of spiral galaxies is largely independent of environment, the intrahalo medium of the host group being the most likely reservoir of cold gas fueling star-formation in satellite galaxies.

Our sample of star-forming galaxies is unlikely to be purely disk-dominated, even if most of the star-formation in the local Universe is shown to occur in disk regions \citep[e.g.][]{James2008}. To exclude the less disky galaxies, we apply an upper limit of 2 to the  $r$-band GALFIT Sersic index of our blue population, which excludes 33\% of the star-forming sample. Transposing the methodology of \citet{Grootes2017} to \umr\ colors, we consider the offset of a galaxy's color from the median value measured for isolated blue (disk) galaxies of the same mass:
\begin{equation}
\Delta(u-r)=(u-r)_{corr} - \overline{(u-r)}^{field}_{corr}(M_\star),
\label{delta_ur}
\end{equation}
where $\overline{(u-r)}^{field}_{corr}(M_\star)$ is a linear fit to the median color-mass relation (Fig.~\ref{cmd}) of blue field galaxies in our mass and redshift ranges. We do not find that this relation varies with large-scale density, nor does the mean color of blue group centrals (Fig.~\ref{environments}).

Figure \ref{grootes_halo} shows the $\Delta(u-r)$ distributions of blue field galaxies, blue satellites in the two regimes of halo mass ($\Log(M_{h}/M_\odot) < $ or $> 12.9$), and blue groups centrals (in small halos by design), with their corresponding median values. Two-sample Kolmogorov-Smirnov tests show that: 
 i) blue group centrals are very similar to blue field galaxies in color-mass relation (two-tailed p-value $p=0.2$); ii) blue satellites in small halos (69.4\%) are also compatible with being drawn from the same distribution ($p=0.04$), and iii) blue satellites in massive halos (30.6\%) differ significantly from blue field galaxies ($p \sim 10^{-9}$), with a red offset $\Delta(u-r) = 0.087 \pm 0.015$ mag. The cut in Sersic index sharpens the distinction between the populations but not applying one   does not alter these conclusions.
 
We propose that the blue satellites of massive red centrals (in the massive halo regime) correspond to the population of disk satellite galaxies in which \citet{Grootes2017} found evidence of suppressed star-formation with respect to field galaxies. Gas fueling must be hindered in these massive halos, in which the fraction of red satellites is also significantly enhanced. However most halos appear not to have a significant impact on the star-formation activity of their blue satellites, in agreement with \citet{Grootes2017} who argue for a change in morphological mix rather than a change in the gas fueling process in groups to explain the well-known color-(local) density relation \citep[e.g.][and references therein]{Zehavi2011}, stating that galaxies are redder in denser environments. We postpone a morphological investigation of group galaxies to a future study.

\section{Caveats}
\label{sec:caveats}


Our analysis is based on a \fof algorithm that was optimized on mock catalogs \citep{Robotham2011} to yield a high group detection rate and a low rate of contamination by interlopers (the best balance between completeness and purity). The linking lengths we selected are in good agreement with the combination found to be optimal for studies of environmental effects by \citet{DuarteMamon2014}. However no group finder is perfect, and it is impossible to recover all the groups exactly as they are known to be in the mocks: some groups will be fragmented and/or merged, contain galaxies unrelated to them, and miss members that will be misplaced in neighboring groups or left alone in the field. Misidentifications of satellites and centrals will follow, and group masses will be wrongly attributed, adding to the uncertainty inherent to any method of assigning a central and a halo mass to each group.

As mentionned in Section \ref{subsec:fofconstruction} and detailed in Appendix \ref{appendix:results}, there is a degeneracy between the two free parameters $b_\perp$ and $R$ used to optimize the cost function, which is why we considered additional statistical quantities, as described in Appendix \ref{appendix:quality}. 
Our choice of parameters, ($b_\perp$,$R$)=(0.06,19), corresponds to the highest purity. However while purity may be most important when comparing satellites to centrals, prioritizing completeness may be preferred to study conformity.   
To try and evaluate the impact of our choice, we tested a combination of parameters (0.07,19.0) that maximises the completeness instead of purity and found no substantial changes in any of our results (conformity, group quenching, star-formation in groups). Additionally, instead of using an optimization based on the global grouping efficiency and global grouping purity, we considered the combination of parameters that maximises the global purity alone (0.07,23.5), and the one that maximizes the completeness alone (0.06,25.5). Again, we found all the results to be qualitatively similar, albeit noisier. 


These different combinations may be considered similar by design as they belong to the zone of degeneracy of the same algorithm and the changes in purity and completeness are relatively small.  However our red fractions of centrals and satellites (Fig.~\ref{redfraction_mass}) are also in qualitative agreement with previous studies using the official SDSS group catalog, and can also be reproduced using the GAMA group catalog of \citet{Robotham2011}, despite significant differences in their rates of fragmentation and/or merging (Section \ref{subsec:fofconstruction}).


\citet{Campbell2015} showed that  accounting for the systematic errors associated with a particular group finder is best achieved by running the group finder over mock catalogs that include realistic galaxy colours and proposed a new statistic (called HTP for halo transition probability) to weigh the combined impact of the above errors. While such forward modeling is beyond the scope of the present paper, their general conclusions shed light on the uncertainties that may plague our results (and others that also do not account for these errors \citep[e.g.][]{Weinmann2006,Yang2008,Yang2009,Peng2012,Wetzel2012,Knobel2015}).

\citet{Campbell2015} created a mock by populating the dark matter halos of a large $N$-body simulation with galaxies of different luminosities and colors ($g-r$), using both subhalo abundance matching and age matching \citep{Hearin2013}, which reproduces both one-halo and two-halo conformity \citep{Hearin2014}. Three different types of group finder were tested on this mock to assess their ability to recover color-dependent halo occupation statistics, including satellite fractions, red fractions and conformity. The resulting group catalogs were found to be remarkably similar (which is reassuring and in agreement with our tests), and to recover most color-dependent statistics reasonably well. In particular, the difference between the red fractions of centrals and satellites at fixed $r$-band luminosity (not shown as a function of stellar mass) is qualitatively reproduced. 

The three group finders tested by \citet{Campbell2015} also recover galactic conformity at fixed halo mass, defined using abundance matching on total group luminosity, but also tend to create weak conformity when this property is removed from the mock. They also found that the strength of the recovered or induced conformity signal at fixed halo mass may be reduced or enhanced, depending on whether luminosity or stellar mass is the primary galaxy property driving halo occupation in the mock data, and in the recovered groups. In other words, the definition of the ``true" and inferred halo masses (which can be both based on luminosity, or on stellar mass, or on one property for the mock and the other for the \fof), have a substantial impact on the systematic uncertainties that are assessed through forward modeling.                  

Nevertheless, we understand from \citet{Campbell2015} that whatever the choice of primary galaxy property, or properties, none of the three \fof algorithms can completely abolish conformity at fixed halo mass when it exists in the mock, nor induce a significant conformity signal when it does not. Running our \fof algorithm on real data, we find that conformity does indeed depend on the definition of halo mass, that it is weak if we use the total group stellar mass, and non existent in a scenario that assigns larger halo masses to red centrals than to blue ones. We conclude from the work of \citet{Campbell2015} that our particular choice of \fof algorithm and parameters is unlikely to be solely responsible for these results. 





\section{Summary and conclusions}
\label{sec:conclusions}

We investigated the properties of central and satellite galaxies in groups at redshift  $z \lesssim 0.2$ and with stellar mass $\Log (M_{\star} / M_\odot) > 10.25$ using the spectroscopic survey Galaxy and Mass Assembly (GAMA). The group catalog was constructed using an anisotropic Friends-of-Friends algorithm taking into account the effects of redshift-space distortion. Red (quiescent) and blue (star-forming) galaxies were classified according to their dust-corrected $(u-r)$ color, shown to be a good bimodal measure of star-forming activity. We explored the fraction of quiescent galaxies (red fraction) in different environments. Our density contrast is defined as the density of central galaxies (satellites were excluded), smoothed by a 3D Gaussian kernel of $\sigma=5$ Mpc and normalized by the redshift-dependent mean density of the survey (using an 8 Mpc scale yields similar results). This estimator probes beyond the virial radius of all groups and is therefore quite different from the ``fifth nearest neighbor" generally used in satellite quenching studies, and which is sensitive to the varying size of DM halos. Our results can be summarised as follows: 


\begin{enumerate}
\itemsep0.3em 

\item \textit{Mass and density quenching:} The red fraction of all galaxies, whether isolated or in groups, is a strongly increasing function of stellar mass, a phenomenon referred to as mass quenching. At fixed stellar mass, the red fraction of all galaxies, including isolated galaxies, increases with the large-scale density contrast. We account for both these effects by defining a quenching efficiency (QE) that separates out both mass and density quenching, designed to be null for field (isolated) galaxies.

\item \textit{Galactic conformity:} The average red fraction of satellites around quenched centrals is significantly higher than that of satellites of the same stellar mass around star-forming centrals. Their QE increases with their central's color, group stellar mass and large-scale density, while that of satellites of blue centrals appears to be independent of all three parameters. This creates a conformity signal that increases with density and group stellar mass. Most of the signal originates from the most massive groups in the densest environments around quenched centrals, in a group stellar mass regime devoid of blue centrals. Some amount of conformity remains at fixed group stellar mass and density. 

\item \textit{Halo quenching:} Assuming group mass traces halo mass, galactic conformity is indeed observed at fixed halo mass as originally claimed. However, if red centrals inhabit more massive halos than blue ones of the same stellar mass, as several studies suggested, and we assume a color-dependent halo-to-stellar-mass ratio, we find that conformity disappears entirely at fixed halo mass. Two quenching regimes emerge: at $\Log(M_{h}/M_\odot)\lesssim13$, centrals still undergo quenching but conformity is insignificant at any given stellar mass and density; at $\Log(M_{h}/M_\odot)\gtrsim13$, a cutoff that sets apart massive ($\Log(M_{\star}/M_\odot)>11$) red only centrals, conformity is meaningless, and the satellites undergo significantly more quenching than their counterparts in smaller halos, all the more so as density increases. This accounts for the conformity signal that increases with density when both regimes are mixed.

\item \textit{Group quenching:} In the low halo mass regime, satellites and group centrals exhibit the same quenching excess, $\sim 10\%$ in red fraction, over field galaxies at fixed stellar mass and density, in agreement with the notion of group quenching advocated by \citet{Knobel2015}, who argued against the importance of satellite-specific processes. However in the high halo mass regime where satellites still undergo quenching while their centrals are fully quenched, the central/satellite dichotomy cannot be ruled out. In this regime, satellite quenching strongly depends on large-scale density, which correlates with halo mass.

\item \textit{Star-formation activity in groups:}  Star-forming group centrals and the majority of star-forming satellites, which reside in low mass halos, show no deviation from the color$-$stellar mass relation of blue field galaxies. However star-forming satellites in high mass halos ($\sim 30\%$) significantly deviate from the color distribution of blue field galaxies, with a mean $(u-r)$ reddening of $\sim + 0.09$ magnitude. 

\end{enumerate}

\section*{Acknowledgements}



We thank our anonymous referee for providing comments that significantly improved this work. 

This research was carried out within the framework of the Spin(e) collaboration (ANR-13-BS05-0005, http://cosmicorigin.org).

GAMA is a joint European-Australasian project based around a spectroscopic campaign using the Anglo-Australian Telescope. The GAMA input catalogue is based on data taken from the Sloan Digital Sky Survey and the UKIRT Infrared Deep Sky Survey. Complementary imaging of the GAMA regions is being obtained by a number of independent survey programmes including GALEX MIS, VST KiDS, VISTA VIKING, WISE, Herschel-ATLAS, GMRT and ASKAP providing UV to radio coverage. GAMA is funded by the STFC (UK), the ARC (Australia), the AAO, and the participating institutions. 
The VISTA VIKING data used in this paper are based on observations made with ESO Telescopes at the La Silla Paranal Observatory under pro- gramme ID 179.A-2004.
The GAMA website is \href{http://www.gama-survey.org/}{http://www.gama-survey.org/}.


\appendix


\section{Group reconstruction Methods}
\label{appendix:method}

It is known that the identification of the groups in redshift space is hampered by several difficulties. Of particular concern are redshift-space distortions such as the elongation of groups along the line of sight due to the peculiar velocities of the group galaxies (the ``fingers-of-God" effect), and the less significant distortions caused by the coherent infall of galaxies towards the center of assembling structures (the ``Kaiser effect"). Both of these effects cause confusion when determining group membership, resulting in excessive merging of galaxies into large structures or, contrarily, the fragmentation of large groups into smaller structures, depending on the adopted strategy for finding the groups.  Without the measure of an absolute distance it is impossible to separate the peculiar velocities from the Hubble flow and these complications can not be fully avoided. 


To surmount these difficulties, different algorithms have been developed, most notably the Friends-of-friends (\fof) algorithm \citep{HuchraGeller1982} recursively creating links between galaxies within a specified volume around each galaxy.
An alternative and more complicated approach to group identification is the Voronoi-Delaunay method developed by  \cite{Marinoni2002} which makes use of the Voronoi-Delaunay tessellation to locally measure clustering parameters on a cluster-to-cluster basis (rather than on a galaxy-to-galaxy basis as in the \fof algorithm). This method was introduced to overcome some of the difficulties inherent to the standard \fof algorithm, but modified \fof implementations have proved competitive 
(e.g., \citealp{Robotham2011} found that while the VDM variant of \cite{Gerke2005} worked reasonably well for larger groups and clusters, their \fof implementation performed better in the low halo mass regime; see also \citealp{Knobel2009} for an extensive performance comparison of both algorithms). In this paper we adopted a \fof algorithm, as described below. 

\section{Friends-of-friends}
\label{appendix:fof}

To deal with the effects of redshift distortions, the distance between two galaxies $i$ and $j$ is measured with two coordinates, the parallel ($d_{\parallel,ij}$) and perpendicular ($d_{\perp,ij}$) projected comoving separations to the mean line-of-sight. Let $\vec{r_i}$ and $\vec{r_j}$ denote the redshift positions of a pair of objects $i$ and $j$,  we define the mean line-of-sight (\los) as : 
\begin{equation}
\label{eq:l}
\vec{l} \equiv \frac{1}{2}\left( \vec{r_i} + \vec{r_j} \right)
\end{equation}
and the redshift space separation is given by :
\begin{equation}
\label{eq:s}
\vec{s} \equiv \vec{r_i} - \vec{r_j}.
\end{equation}
 
The projected parallel and perpendicular line-of-sight separations between $i$ and $j$, $d_{\perp,ij}$ and $d_{\parallel,ij}$ respectively, are then given by :
\begin{equation}
\label{eq:d_parallel}
d_{\parallel,ij} = \frac{\vec{s} \cdot \vec{l}}{\parallel \vec{l} \parallel}
\end{equation}
and
\begin{equation}
\label{eq:d_perp}
d_{\perp,ij} = \sqrt{\vec{s} \cdot \vec{s} - d_{\parallel,ij}^2}.
\end{equation}

The two galaxies are linked to each other if : 
\begin{equation}
\label{eq:link_perp}
d_{\perp,ij} < b_\perp \overline{r}_{ij} 
\end{equation}
and
\begin{equation}
\label{eq:link_parallel}
d_{\parallel,ij} < b_\parallel \overline{r}_{ij}.
\end{equation}
Here $b_\perp$ and $b_\parallel$ are the projected and line-of-sight linking lengths 
in units of  the mean intergalactic separation given by :
\begin{equation}
\label{eq:mean_sep}
\overline{r}_{ij} = \frac{1}{2}\left(n_i^{-1/3} + n_j^{-1/3}\right),
\end{equation}
with $n_i$ and $n_j$ being the galaxy number densities at the redshifts of galaxies $i$ and $j$. 
The parameter $b_\parallel$ is related to $b_\perp$ through the radial expansion factor $R = b_\parallel / b_\perp$ accounting for the peculiar motions of galaxies within groups. 



\section{Parameters optimization strategy}
\label{appendix:optimization}

The \fof algorithm described above has two free parameters, the linking length $b_\perp$ and the radial expansion factor $R$ (or equivalently the perpendicular and line-of-sight linking lengths). Their values will affect the quality of the resulting group catalog: too small values will tend to break up single groups into several groups, while too large values will merge multiple groups into single ones. 

These free parameters can be determined by optimizing a group cost function (a measure of the grouping quality) when tested on the mock catalogs. 
No combination of linking lengths will create a group catalog recovering simultaneously all aspects of the underlying halo distribution, no matter how large the systems are \citep[see e.g.][]{Berlind2006}. 
The optimization strategy depends on the scientific purpose of the group catalog. 
Our objective is to obtain a catalog with a high group detection rate and a low contamination by galaxies coming from different groups. We followed the definition of the group cost function of \cite{Robotham2011}, with slightly different notations and minor modifications.


We need to define the way reconstructed groups (the \fof groups) are associated with the underlying real groups in the mock catalogs (the mock groups). Ideally, the associations between the real and reconstructed groups should be bijective, meaning that the joint galaxy population of the \fof and mock groups includes more than 50 \% of their respective members. Such an association is unambiguous, as each group can bijectively match one group at most, and the reconstructed group catalog and the corresponding real group catalog are mutually an accurate representation of each other. 

To cast such a two-way grouping quality into a statistical measure, let $\Ngbij$, $\Ngfof$ and $\Ngmock$ denote the number of bijective, \fof and mock groups, respectively.
Following~\cite{Robotham2011}, we define :

\begin{equation}
\Efof = \frac{\Ngbij}{\Ngfof},
\end{equation}

\begin{equation}
\Emock = \frac{\Ngbij}{\Ngmock},
\end{equation}
and

\begin{equation}
\label{Eq:etot}
\Etot = \Efof \Emock.
\end{equation}
By definition, these measures take values in the range 0-1. $\Etot$ can be seen as the global halo finding efficiency, telling us how well the groups are recovered. It will be 1 if all groups are bijective, and 0 if there is no group with a bijective match. 

Another measure of the grouping quality is how well galaxies within the groups are recovered. Let $\Pfofij$ and $\Pmockij$ denote the purity products 
defined as follows :
\begin{equation}
\Pfofij  =  \frac{N_{\mathrm{m}_{\mathrm{FoF},i \cap \mathrm{mock},j}} }{\Nmfofi} \frac{N_{\mathrm{m}_{\mathrm{FoF},i \cap \mathrm{mock},j}} }{\Nmmockj},
\end{equation}
where $N_{\mathrm{m}_{\mathrm{FoF},i \cap \mathrm{mock},j}}$ is number of galaxies the $i^{th}$ FoF group shares with the $j^{th}$ mock group, and
\begin{equation}
\Pmockij  = \frac{N_{\mathrm{m}_{\mathrm{mock},i \cap \mathrm{FoF},j}} }{\Nmmocki}  \frac{N_{\mathrm{m}_{\mathrm{mock},i \cap \mathrm{FoF},j}} }{\Nmfofj}, 
\end{equation}
where $N_{\mathrm{m}_{\mathrm{mock},i \cap \mathrm{FoF},j}}$ is the number of galaxies that the $i^{th}$ mock group shares with the $j^{th}$ \fof group. 
We will call the best match the association between the $i^{th}$ mock (\fof) group and the $j^{th}$ \fof (mock) group for which $\Pmockij$ ($\Pfofij$) is highest. 


We then define the following measures :
\begin{equation}
\Qfof =  \frac{ \sum_{i=1}^{\Ngfof}  N_{\mathrm{m}_{\mathrm{FoF},i \cap \mathrm{mock},j}} \Qfofi}{\sum_{i=1}^{\Ngfof} \Nmfofi },
\end{equation}
where
\begin{equation}
\Qfofi =  \frac{  N_{\mathrm{m}_{\mathrm{FoF},i \cap \mathrm{mock},j}}}{ \Nmfofi },
\end{equation}
and
\begin{equation}
\Qmock = \frac{ \sum_{i=1}^{\Ngmock}  N_{\mathrm{m}_{\mathrm{mock},i \cap \mathrm{FoF},j} \Qmocki} }{\sum_{i=1}^{\Ngmock} \Nmmocki},
\end{equation}
where
\begin{equation}
\Qmocki =  \frac{  N_{\mathrm{m}_{\mathrm{mock},i \cap \mathrm{FoF},j}}}{ \Nmmocki }.
\end{equation}
$\Nmfofi$ and $\Nmmocki$ denote the number of members in the $i^{th}$ \fof and mock 
group, respectively, and $N_{\mathrm{m}_{\mathrm{FoF},i \cap \mathrm{mock},j}}$ ($N_{\mathrm{m}_{\mathrm{mock},i \cap \mathrm{FoF},j}}$) the number of $i^{th}$ \fof (mock) group members recovered by its corresponding matched mock (\fof) group. 



Finally the global grouping purity is defined as :
\begin{equation}
\label{Eq:qtot}
\Qtot = \Qfof \Qmock.
\end{equation}

$\Qtot$ will be 1 if all galaxies within all groups are perfectly recovered, its lower limit will be however greater than 0 as it is always possible to break a catalog with $N$ galaxies into a catalog of N groups. 

The final quantity defining our group cost function is :
\begin{equation}
\label{Eq:stot}
\Stot = \Etot \Qtot,
\end{equation}
which takes values between 0 and 1 and should be maximised.

In order to find the right combination of linking lengths, we use the publicly available GAMA mock catalogs\footnote{ \url{http://www.gama-survey.org} } described below, on which we run our \fof group-finder algorithm for a grid of $b_\perp$ and $R$ values, and compute $\Stot$. 

\section{Mock catalogs}
\label{appendix:mocks}

The GAMA mock galaxy catalogs (see  \cite{Robotham2011} for a detailed description) were constructed from the Millennium dark matter N-body simulation \citep{Springel2005} run with the following cosmological parameters: $\Omega_m = 0.25$, $\Omega_{\Lambda} = 0.75$, $\Omega_b = 0.045$, $h = 0.73$, $n = 1$ and $\sigma_8 = 0.9$.
The DM halos were then populated with galaxies using the \cite{Bower2006} variant of the GALFORM semi-analytic model of galaxy formation. 

There are nine mock catalogs, each of which covers a complete analog of the full GAMA I survey, i.e. three regions of 12 $\times$ 4 deg$^2$ out to redshift 0.5, while preserving the true angular separation between them. 
The public mocks are limited to the apparent magnitude $r < 19.4$ but as shown by \cite{Robotham2011}, there is no difference in the resulting optimised parameters to $r=19.8$.
Since the nine mock catalogs are constructed form a single simulation, they are not statistically independent. In spite of this limitation, the construction method guarantees that no spatial overlap between the nine light cone mocks is present and there is no single galaxy at the exact same stage of evolution in more than one mock.

\section{Results of the optimization}
\label{appendix:results}

As we already mentioned, the optimization of the group-finding parameters should be carried out in a way that the resulting group catalog best fulfils the desired scientific goals. For our optimization we use only groups with five or more members which are allowed to match groups with two or more members. This choice is motivated by the desire to estimate the global group properties of the resulting group catalog. 

The results of the optimization computation are shown in Fig.~\ref{Fig:stot}. The global maximum of $\Stot$ is obtained for the combination of two parameters ($b_\perp$,$R$) = (0.06,27.5). We note however, that between $R \gtrsim 16$ and $R=30$, $\Stot$ does not evolve significantly. This remains true for $b_\perp = 0.07$ as well. In addition, the values of $\Stot$ for these two linking lengths, and in the $R$ range 16-30 are very similar within the error bars. This means that increasing the value of $R$ in a given range leads to very similar global statistical properties of the reconstructed catalogs. We will thus include an additional criterion 
by considering the contribution from the mock and \fof components to the overall cost function. This contribution should be as symmetric as possible, indicating that the \fof algorithm recovers on average similar groups, in terms of number and quality of reconstruction, that actually exist in the mock catalog. 
Table~\ref{tab:opt} shows the \fof parameters corresponding to the maximum global cost function $\Stot$, and 
the most symmetric contribution to $\Stot$ from the mock and \fof components ($\Smock = \Emock \Qmock$ and $\Sfof = \Efof \Qfof$, respectively). For the global maximum of the cost function, corresponding to ($b_\perp$, $R$) = (0.06, 27.5), the cost from mock groups to $\Stot$ is lower than the one from \fof groups indicating that the group-finding algorithm globally recovers less groups than actually exist in the mock catalog. Similar asymmetry is found for the combination ($b_\perp$, $R$) = (0.07, 19.0), corresponding to the maximum of $\Stot$ for $b_\perp$ = 0.07, and ($b_\perp$, $R$) = (0.07, 15.0), which is the most symmetric solution for $b_\perp$ = 0.07. The most equilibrated contribution is obtained for ($b_\perp$, $R$) = (0.06, 19.0), which is our preferred choice for the \fof parameters.

\begin{table}
\centering
\caption{The \fof parameters.\label{tab:opt}}
\begin{tabular}{cccccc}
\hline \hline
&$b_\perp$ & $R$ & $\Stot$ & ${\Smock}^e$ & ${\Sfof}^f$\\
\hline \hline
A$^a$ & 0.06 & 27.5 & 0.405  & 0.602 & 0.673  \\
B$^b$ & 0.07 & 19.0 & 0.393  & 0.556 & 0.707  \\
C$^c$ & 0.06 & 19.0 & 0.381  & 0.618 & 0.617  \\
D$^d$ & 0.07 & 15.0 &  0.381 & 0.568 &  0.671 \\
\hline
\end{tabular}
\begin{description}
\itemsep0em 
\item $^{a}${\small global maximum of $\Stot$}\\
\item $^{b}${\small maximum $\Stot$ for $b_\perp$ = 0.07}\\
\item $^{c}${\small the most symmetric contribution to the $\Stot$ from the mock and \fof components for $b_\perp$ = 0.06} \\
\item $^{d}${\small the most symmetric contribution to the $\Stot$ from the mock and \fof components for $b_\perp$ = 0.07}\\
\item $^{e}${\small contribution to the $\Stot$ from the mock component}\\
\item $^{f}${\small contribution to the $\Stot$ from the \fof component}
\end{description}
\end{table}

\section{Quality of group reconstruction}
\label{appendix:quality}

The statistical measures used to define our cost function introduced in the previous Section already allow us to assess the performance of our group-finding algorithm. We will however introduce additional statistical quantities, different from those used in the optimization, in order to test the quality of the reconstructed catalog independently. Following~\cite{Knobel2009}\footnote{The statistical quantities defined in this Section follow with some modifications those introduced in~\cite{Knobel2009}. Some of these measures can be found in their original form in~\cite{Gerke2005}.}, we define two classes of statistical quantities. One on a group-to-group basis, including the completeness, purity, overmerging and fragmentation, and the second one on a galaxy-to-group basis, with the galaxy success rate and the interloper fraction. We will define each of these quantities for the best and bijective matches as introduced in Section~\ref{appendix:optimization}. 

Let $\Nbestgmock(\geq N)$ and $\Nbijgmock(\geq N)$ be the number of mock groups with $N$ or more members for the best and bijective matches respectively\footnote{In all the following definitions in this section, if not stated differently, the richness $N$ refers to the richness of groups that are being matched (mock groups in this case). In the matching procedure, the minimum richness of groups in the reference sample (\fof groups in this case) is two.}

We define the completeness $\cbest(N)$ and $\cbij(N)$ as the fraction of real groups with $N$ or more members that are successfully recovered in the reconstructed group catalog corresponding to the best and bijective match, respectively:

\begin{equation}
\label{Eq:cbest}
\cbest(N) = \frac{\Nbestgmock(\geq N)}{\Ngmock(\geq N)}, 
\end{equation}

\begin{equation}
\label{Eq:cbij}
\cbij(N) = \frac{\Nbijgmock(\geq N)}{\Ngmock(\geq N)}.
\end{equation}

Similarly, if $\Nbestgfof(\geq N)$ and $\Nbijgfof(\geq N)$ denote the number of \fof groups with $N$ or more members in the best and bijective matches respectively, the purity $\pbest(N)$ and $\pbij(N)$ are defined as the fractions of reconstructed groups with $N$ or more members belonging to real groups:

\begin{equation}
\label{Eq:pbest}
\pbest(N) = \frac{\Nbestgfof(\geq N)}{\Ngfof(\geq N)},
\end{equation}

\begin{equation}
\label{Eq:pbij}
\pbij(N) = \frac{\Nbijgfof(\geq N)}{\Ngfof(\geq N)}.
\end{equation}

By definition, the quantities $\cbest$, $\cbij$, $\pbest$ and $\pbij$ take values between 0 and 1. The completeness is one if all real groups are reconstructed while it is 0 if no real group is detected. Similarly, the purity is 1 if all reconstructed groups are matched with real groups, 0 if all reconstructed group are spurious (none is associated with any real group). 

For each mock catalog, we define fragmentation as the inverse of the number of \fof groups it is matched with. Similarly, for each \fof group overmerging is given as the inverse of the number of mock groups it is matched with. These quantities are well defined only for groups that have been actually matched (bijective or best match). Fragmentation (overmerging) is equal to one if each matched mock (\fof) group is associated with only one \fof (mock) group and the smaller the values are, the more \fof (mock) groups are associated with a given mock (\fof) group.

The galaxy success rate  $\Sgal(N)$ is defined as the fraction of galaxies in real groups with $N$ or more members that are found to belong to any reconstructed group with 2 or more members. For best and bijective matches, these definitions become :

\begin{equation}
\label{Eq:Sgalbest}
\Sgalbest = \frac{  N^{\mathrm{best}}_{\mathrm{m}_{\mathrm{mock} \cap \mathrm{FoF}}}(\geq N)}{ \Nmmock(\geq N) }.
\end{equation}
and
\begin{equation}
\label{Eq:Sgalbij}
\Sgalbij = \frac{  N^{\mathrm{bij}}_{\mathrm{m}_{\mathrm{mock} \cap \mathrm{FoF}}}(\geq N)}{ \Nmmock(\geq N) },
\end{equation}
respectively.


We define the interloper fraction $\finter(N)$ as the fraction of galaxies in reconstructed groups having $N$ or more members that do not belong to any matched real group of richness $\geq 2$.
We again distinguish best and bijective match:

\begin{equation}
\label{Eq:fibest}
\finterbest = \frac{  \Nmfofbest(\geq N)  - N^{\mathrm{best}}_{\mathrm{m}_{\mathrm{mock} \cap \mathrm{FoF}}}(\geq N) }{ \Nmfof(\geq N) },
\end{equation}
\begin{equation}
\label{Eq:fibij}
\finterbij = \frac{  \Nmfofbij(\geq N)  - N^{\mathrm{bij}}_{\mathrm{m}_{\mathrm{mock} \cap \mathrm{FoF}}}(\geq N) }{ \Nmfof(\geq N) }.
\end{equation}
In the above definitions, an interloper is defined as any galaxy in a reconstructed group not belonging to its matched real group. We also define interlopers as field galaxies that end up in reconstructed groups. The corresponding interloper fraction $\finterfield(N)$ is the fraction of galaxies belonging to reconstructed groups with $N$ or more galaxies that are field galaxies.
%
 $\Sgal$, $\finter$ and $\finterfield$ take values in the range 0-1. By definition, $\finterfield$ is lower or equal to $\finter$ (equal if all interlopers are field galaxies).

\begin{figure}
\begin{center}
\includegraphics[width=\columnwidth]{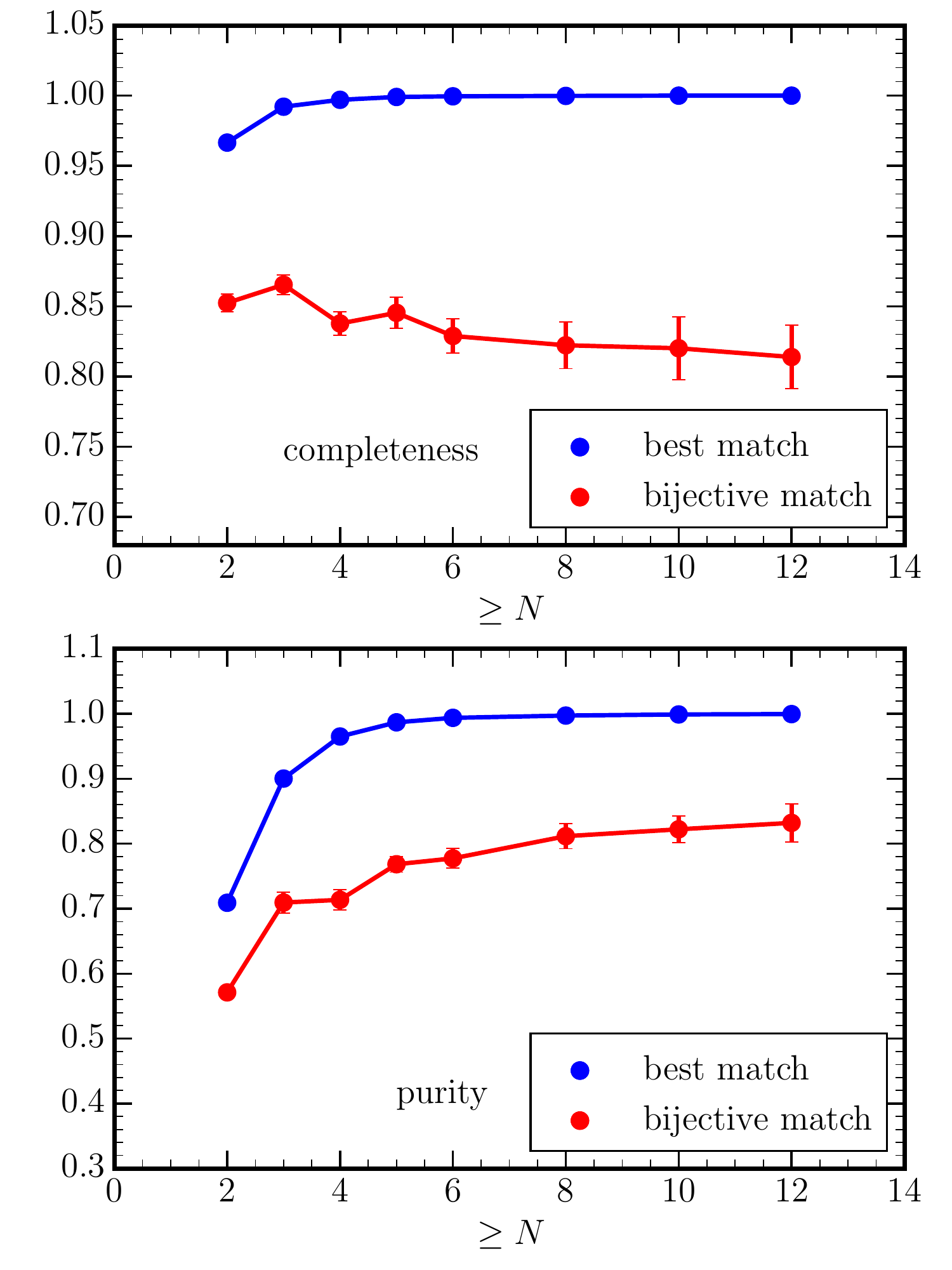}
\caption{\label{Fig:c_p_vs_N} Completeness (upper panel) and purity (lower panel) as a function of richness $N$ for the best (blue) and bijective (red) match. The points represent the mean values among the 9 mock catalogs and the error bars show their scatter. 
}
\end{center}
\end{figure}

\begin{figure}
\begin{center}
\includegraphics[width=\columnwidth]{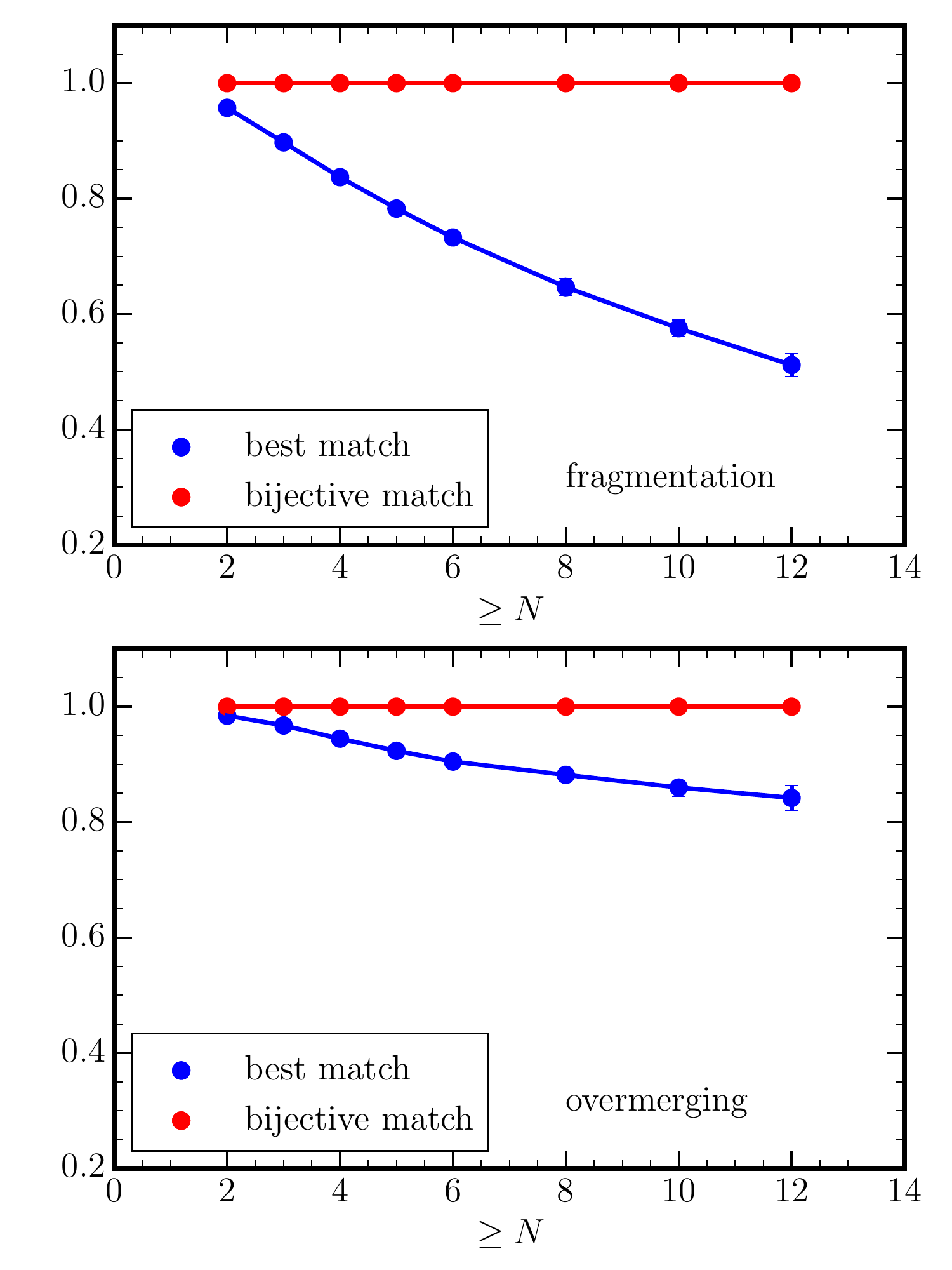}
\caption{\label{Fig:frag_over} Fragmentation (upper panel) and overmerging (lower panel) as a function of richness $N$ for the best (blue) and bijective (red) match.}
\end{center}
\end{figure}

\begin{figure}
\begin{center}
\includegraphics[width=\columnwidth]{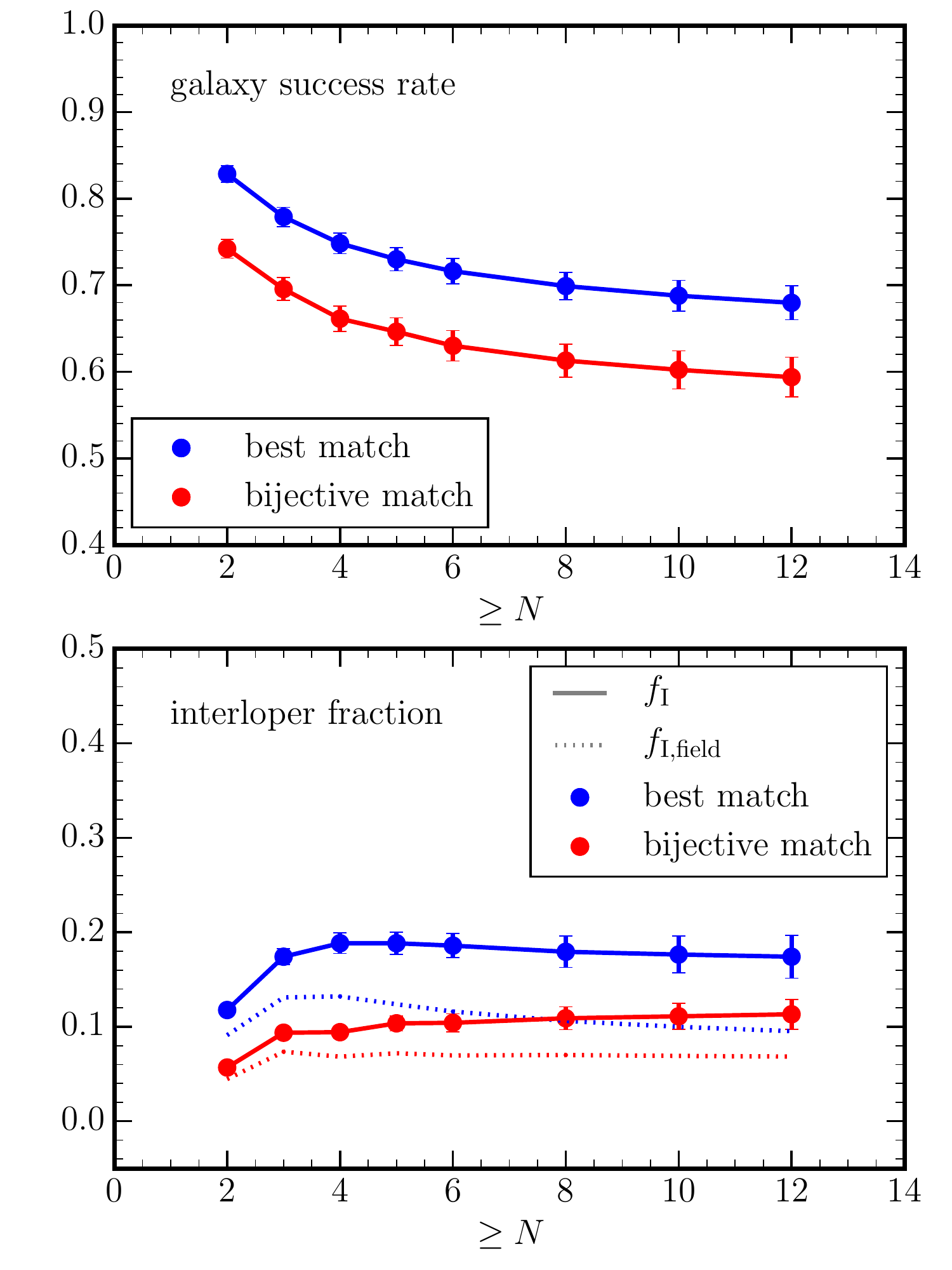}
\caption{\label{Fig:gal_success} Galaxy success rate (upper panel) and interloper fraction (lower panel) as a function of richness $N$ for the best (blue) and bijective (red) match. In the lower panel, the solid line corresponds to the interloper fraction as defined by Equations~\ref{Eq:fibest} and \ref{Eq:fibij}, and the dashed line refers to the definition of interlopers as field galaxies ($\finterfield$). For the sake of clarity in the lower panel, error bars are shown only for $\finter$.}
\end{center}
\end{figure}

\begin{figure}
\includegraphics[width=\columnwidth]{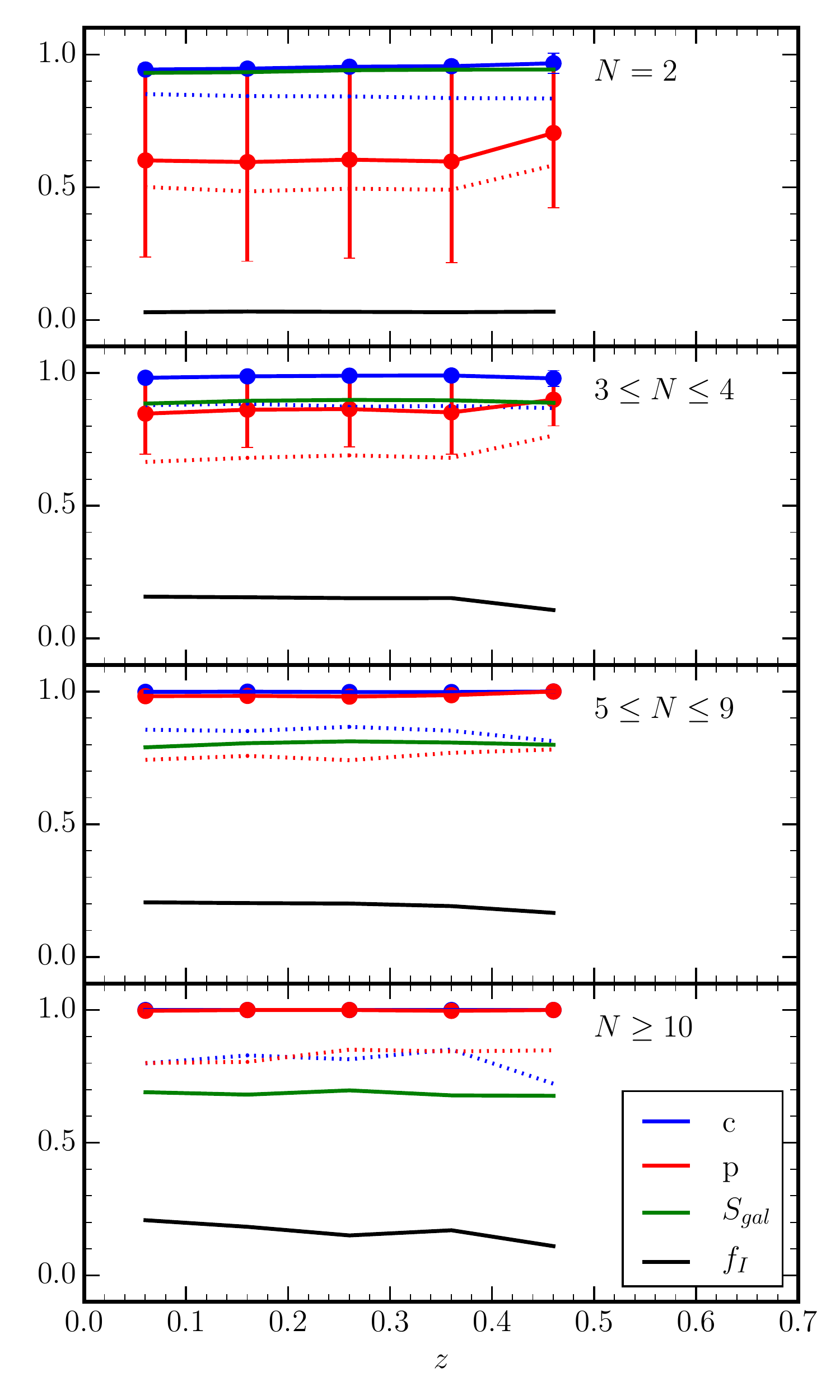}
\caption{\label{Fig:all_vs_z_N} Completeness (blue), purity (red), galaxy success rate (green) and interloper fraction (black) as a function of redshift for different bins of $N$. The solid and dotted lines correspond to the best and bijective matches, respectively. For the sake of clarity, error bars and bijective matches are shown only for completeness and purity.}
\end{figure}

All of the above defined quantities can be combined into three statistical measures of quality 
of our group reconstruction as follows :
\begin{equation}
\label{eq:g1}
g_1 = \sqrt{(1-\cbest)^2+(1-\pbest)^2},
\end{equation}

\begin{equation}
\label{eq:g2}
g_2 =  \mathrm{overmerging}^{\mathrm{best}} \times \mathrm{fragmentation}^{\mathrm{best}}
\end{equation}
and
\begin{equation}
\label{eq:g3}
g_3 = \sqrt{(1-\Sgalbest)^2+(\finterbest)^2}.
\end{equation}
For a perfectly reconstructed group catalog $\cbest$, $\pbest$, overmerging, fragmentation and $\Sgal$ are equal to one, and $\finter = 0$. As the reconstruction of such a perfect catalog is not possible, the best we can do is to try and find a balance between different quantities that often tend to be exclusive. This is the case of completeness and purity, the balance of which is measured by $g_1$. Increasing the completeness of a group catalog often results in decreasing its purity (less undetected groups, more spurious ones). Similarly, $g_3$ is a measures of balance between spurious groups and interlopers. $g_1$ and $g_3$ should thus be minimised.   
A good group catalog should also avoid overmerging and fragmentation, so $g_2$ should be maximised.  

We recall that the purpose of the above defined measures is not the optimization of the group-finder parameters, but rather an attempt to break the 
degeneracy between several combinations of potentially optimal parameters. A summary of the statistics computed for $N\geq 5$ for such combinations is shown in Table~\ref{tab:stats}. We notice that there is no couple of linking lengths for which all three goodness parameters are optimised at the same time. Based on the argument of symmetry between real and reconstructed group catalogs, we selected in Section~\ref{appendix:optimization} the combination of linking lengths ($b_\perp$, $R$) = (0.06, 19.0) as being optimal. This set of parameters leads to the highest balance between completeness and purity (lowest $g_1$), and lowest interloper fraction among all four considered combinations. The \fof parameters ($b_\perp$, $R$) = (0.07, 19.0) result in a catalog with the lowest overmerging and fragmentation (highest $g_2$), and lowest number of spurious groups (highest $\Sgal$), again among the four selected combinations. However, they also produce the highest asymmetry between the mock and \fof counterparts. Thus it seems reasonable to keep ($b_\perp$, $R$) = (0.06, 19.0) as our best choice of optimal \fof parameters for the construction of our group catalog. 

\begin{table}
\centering
\caption{Summary of statistics for $N \geq 5$ (same combinations of \fof parameters as in Table~\ref{tab:opt}). 
\label{tab:stats}
}
\begin{tabular}{cccccccc}
\hline \hline
  &$b_\perp$ & $R$ & $g_1$ & $g_2$ & $g_3$ & $\Sgal$ & $\finter$ \\
\hline \hline
A & 0.06 & 27.5 & 0.016  & 0.724 & 0.308 & 0.772 & 0.208 \\
B & 0.07 & 19.0 & 0.024  & 0.739 & 0.306 & 0.808 & 0.238 \\
C & 0.06 & 19.0 & 0.013  & 0.723 & 0.329 & 0.729 & 0.188 \\
D & 0.07 & 15.0 & 0.021  & 0.733 & 0.318 & 0.776 & 0.226 \\
\hline
\end{tabular}
\end{table}

The following figures illustrate the quality level of this particular reconstructed catalog. Figure~\ref{Fig:c_p_vs_N} shows the completeness and purity for the best and bijective match as a function of richness $N$. The completeness for the best match $\cbest$ is about 0.99, showing no dependence on the richness $N$, while the completeness for the bijective match $\cbij \simeq 0.84$ only shows a weak dependence on $N$. The purity parameters, $\pbest \simeq 0.94$ and $\pbij \simeq 0.76$, are also almost independent on richness for $N \geq 3$. For $N=2$ both purities drop significantly (to $\simeq 0.7$ for $\pbest$ and $\simeq 0.56$ for $\pbij$). 

Fragmentation and overmerging are shown in Fig.~\ref{Fig:frag_over}. Bijectively matched groups are neither overmerged nor fragmented no matter their richness. For groups matched according to our best criterion, fragmentation is more severe than overmerging. The overall average fragmentation is $\simeq 0.74$, while this value drops to $\simeq 0.65$ for groups with $N \geq 5$. Overmerging decreases with $N$, however only mildly with an average value of $\simeq 0.91$.   

The galaxy success rate and interloper fraction, two statistical quantities on a galaxy-to-group basis, are shown in Fig.~\ref{Fig:gal_success}. For both best and bijective matches these measures show similar dependences on $N$. The galaxy success rates $\Sgalbest \simeq 0.73$ and $\Sgalbij \simeq 0.65$, both decline until $N \simeq 5$, after which the dependence on $N$ weakens. The interloper fraction is overall very low ($\finterbest \simeq 0.17$ and $\finterbij \simeq 0.09$) with almost no dependence on $N$, except for $N = 2$, where it reaches its minimum value.

Finally, Fig.~\ref{Fig:all_vs_z_N} shows the statistics as a function of redshift for different bins of richness $N$. All quantities are consistent with no or very weak evolution with redshift. Only in the highest redshift bin do we note an increase in purity for groups with $N \leq 4$, and a decrease in completeness and interloper fraction for bijective matches among groups of richness $\geq 10$.

\bibliographystyle{aasjournal}
\bibliography{gama_groups}

\end{document}